\documentclass[12pt]{article}
\usepackage{epsfig,youngtab,pdflscape,amssymb}
\usepackage{fancybox,amsmath,color,float}

\def\mC{ \mathbb{C}}  
\def\Sym{ \hbox{Sym} } 
\def\cI{ \mathcal{ I } } 

\def\cA{ \mathcal{A}}

\def\cO{ \mathcal{O} } 
\def\cR{ \mathcal{R} } 
 
\def\cN{ \mathcal{N}}
\def\cH{ \mathcal{H}}

\def\mV{ \mathbb{V} }

\def\cC{ \mathcal{ C } } 
\def\cM{ \mathcal{M}} 

\def\cW{ \mathcal{W}} 
\def\Dim{ \hbox{Dim}} 
\def\cL{ \mathcal{L}}
\def\cP{ \mathcal{P} }
\def\mR{ \mathbb{R}}  

\def\nat{ {\rm{nat}}} 

\newcommand{\be}{\begin{equation}}
\newcommand{\bea}{\begin{eqnarray}}

\newcommand{\ee}{\end{equation}}
\newcommand{\eea}{\end{eqnarray}}

\newcommand{\Mult}{ {\rm Mult} }

\begin{document}

\makeatletter
\@addtoreset{equation}{section}
\makeatother
\renewcommand{\theequation}{\thesection.\arabic{equation}}

\rightline{QMUL-PH-18-06}
\vspace{0.8truecm}

\vspace{10pt}


{\LARGE{ 
\centerline{\bf Free field  primaries  in general dimensions:   } 
\centerline{ \bf  Counting and construction  with rings and modules } 
}}  

\vskip.5cm 

\thispagestyle{empty} \centerline{
    {\large \bf Robert de Mello Koch${}^{a,b,}$\footnote{ {\tt robert@neo.phys.wits.ac.za}}}
   {\large \bf and Sanjaye Ramgoolam
               ${}^{b,c,}$\footnote{ {\tt s.ramgoolam@qmul.ac.uk}}   }
                                                       }

\vspace{.2cm}
\centerline{{\it ${}^a$ School of Physics and Telecommunication Engineering},}
\centerline{{ \it South China Normal University, Guangzhou 510006, China}}

\vspace{.2cm}
\centerline{{\it ${}^b$ National Institute for Theoretical Physics,}}
\centerline{{\it School of Physics and Mandelstam Institute for Theoretical Physics,}}
\centerline{{\it University of the Witwatersrand, Wits, 2050, } }
\centerline{{\it South Africa } }

\vspace{.2cm}
\centerline{{\it ${}^c$ Centre for Research in String Theory, School of Physics and Astronomy},}
\centerline{{ \it Queen Mary University of London},} \centerline{{\it
    Mile End Road, London E1 4NS, UK}}

\vskip.8cm

\thispagestyle{empty}

\centerline{\bf ABSTRACT}

\vskip.2cm 

We define lowest weight polynomials (LWPs), motivated by $so(d,2)$ representation theory, 
as elements of the polynomial ring over $ d \times n $ variables obeying a system of  first and second order partial differential equations. LWPs invariant under $S_n$ correspond to primary fields in free scalar field theory in $d$ dimensions, constructed from $n$ fields. The LWPs are in one-to-one correspondence with  a quotient of the polynomial ring in $ d \times (n-1) $  variables by  an ideal generated by $n$ quadratic polynomials. The implications of this description for the counting and construction of primary fields are described: an interesting binomial identity underlies one of the construction algorithms.The product on the ring of LWPs can be described as a commutative star product. The quadratic algebra of lowest weight polynomials has a dual quadratic algebra which is non-commutative. We discuss the possible physical implications of this dual algebra.

\setcounter{page}{0}
\setcounter{tocdepth}{2}

\newpage

\tableofcontents

\setcounter{footnote}{0}
\linespread{1.1}
\parskip 4pt

{}~
{}~

\section{ Introduction} 

The counting and construction of primary fields in free scalar field theories
was found to have surprisingly simple and elegant geometrical structures in \cite{DRRR17-PRL,DRRR17}. 
General primary fields in scalar field theory in $d$ dimensions, which are composites of 
$n$ elementary fields, are in 1-1 correspondence with 
polynomials in $ n d $ variables, $x_{\mu}^I$ where $ 1 \le \mu \le d, 1 \le I \le n $  which solve 
a system of linear first and second order partial differential equations, and obey 
an invariance condition under $S_n$, the symmetric group of permutations of $n$ distinct objects. 
A holomorphic sector of primaries corresponds to the ring of functions 
on an $S_n$ orbifold. In \cite{HLMM17} it was observed that the space of all primary fields in a scalar theory  corresponds to a quotient ring, and that this ring also arises in the classification of 
effective actions. 

At the core of these developments is a simple problem  in the representation 
theory of the $d$-dimensional conformal algebra $so(d,2)$ and its surprisingly rich connections to polynomial rings, modules over these rings, the standard  mathematics of algebraic geometry, 
 as well as to non-commutative  algebras and their quotients. 
According to the operator-state correspondence in conformal field theory, local operators are in 1-1 correspondence with quantum states. Corresponding to an elementary scalar field in $ d$ dimensions and 
 its non-vanishing 
derivatives is an irreducible representation $V$ of  $so(d,2)$. 
The problem is to decompose 
the tensor product $V^{ \otimes n }$ into irreducible representations of 
$so(d,2)\times S_n$.  A convenient realization of the representation $ V$ is in terms of polynomials 
in $ x_{ \mu}$ while for $V^{ \otimes n } $ we have polynomials in $ x_{\mu}^I$. 
This problem can be approached in two steps : Find all the states in $V^{ \otimes n } $ annihilated by the 
special conformal transformation generators $K_{ \mu}$, then decompose these states 
according to representations of $so(d) \times S_n$. Further projecting to the trivial representation of $S_n$ gives the primary fields for free scalar field theory. The states annihilated by 
$K_{ \mu} $ are the states of lowest conformal dimension in irreducible representations of $so(d,2)$, which may have non-trivial $so(d)$ transformation properties. With the polynomial realization of 
$V^{ \otimes n }$ in hand, these states are certain polynomials in $ \mC [ x_{ \mu}^I ] $, which we call 
{\it lowest weight polynomials } (LWPs). Following   \cite{DRRR17-PRL,DRRR17} we review the fact that LWPs in  $ \mC [ x_{ \mu}^I ] $ are solutions of a system of first and second order 
partial differential equations. 
We explain the 1-1 correspondence between the polynomials and the elements of a quotient ring defined 
in \cite{HLMM17}. The first order equations take the form of a condition of vanishing centre of mass momentum. They can be solved explicitly, leading to a description as a polynomial ring in $ (n-1)d$ variables $ X_{ \mu}^{ A}$, with $ 1 \le A \le (n-1)$ and transforming in the irreducible 
representation of $S_n$ corresponding to the hook-shaped Young diagram $[n-1,1]$ with 
first row of length $n-1$ and second row of length $1$. We will denote this irrep as $V_H$ in the following. 
 Our first new result (Section \ref{sec:quadrels})  is to give  an  explicit description of the quotient   ring in dimension $d$ in terms of $(n-1)d$ generators  and explicit quadratic relations. The quadratic relations are given in terms of a Clebsch-Gordan  decomposition problem for $S_n$, which we explicitly solve. 

The Hilbert series of the quotient ring, which can be deduced from the character of $ V^{ \otimes n} $
implies counting formulae for the number of linearly independent LWPs at each degree in $ \mC [ X_{ \mu}^A]$.  These dimensions are expressed as  an alternating sum of positive quantities.
A transform, which we dub the {\it confluent binomial transform},  
is found which gives the dimensions as sums of positive quantities (Section \ref{sec:CBT}). 
This leads directly to a construction algorithm for the lowest weight polynomials, which we refer to as the first construction algorithm. This is our second main result. We compare this with two additional construction algorithms. Construction II works directly in $ \mC [ x_{ \mu}^I ] $ and imposes 
first and second order conditions. Construction III exploits the quadratic constraints 
and looks at an intersection of projectors.  It exploits analogies between the construction of LWPs
and the construction of traceless tensors of $so(k)$, and as such it 
has links to Brauer algebras which arise 
as commutants of $so(k)$ in tensor spaces.

The paper is organized as follows. Section 2 starts with a review of \cite{DRRR17-PRL,DRRR17,HLMM17}. 
We describe the system of first and second order partial differential equations 
for polynomials in $ \mC [x_{\mu}^I ] $ which define the LWPs. We  explain the correspondence 
with a quotient ring obtained by quotienting out an ideal $\cI$ generated by linear and quadratic constraints. 
We then establish a description of LWPs where we have solved the first order constraints. 
This leads to a  quotient ring of  $ \mC [ X_{ \mu}^A ] $ by quadratic relations. 
These quadratic relations are given explicitly in terms of Clebsch-Gordan coefficients 
for $V_{ H } \otimes V_H$. 

In Section \ref{sec:HS} we use the $so(d,2)$  character of $ V^{ \otimes n} $ to arrive at the 
Hilbert series of the ring $ \cL  $ of lowest weight primaries. Section \ref{sec:exactsequences}
explains the exact sequences of modules over the polynomial ring $ \cR = \mC [ X_{ \mu}^A ] $, which give a resolution of $ \cL$. This exact sequence implies the Hilbert series. 
It also leads directly to exact sequences of vector spaces over $ \mC$.
Section \ref{sec:refinedcounting} extracts counting formulae for LWPs, 
refined according to $so(d) \times S_n$ irreps, which follow from the exact sequences. 

With an understanding of refined counting formulae, we expect to deduce  
algorithms for construction of LWPs. One tricky point is that the counting 
formulae in Section \ref{sec:refinedcounting} involve alternating sums of dimensions
of vector spaces. In section \ref{sec:CBT} 
we show that the counting formula for dimensions of LWPs 
obtained from the Hilbert series is equivalent to a formula as a sum of positive 
constructible quantities. An important feature is that this constructive formula at fixed degree $k$ 
is expressed in terms of LWPs at lower degrees. Both the positive formula and the alternating sum formula 
involve binomial coefficients. If we denote by $V_Q$ the vector space of quadratic constraints, 
the alternating sum formula involves dimensions of  exterior powers of  $V_Q$, while the positive formula 
involves dimensions of symmetric powers of $V_Q$. The key identity responsible for this inversion relating the positive and alternating sum formula (\ref{SBT}) turns out to be a special value of 
a confluent hypergeometric function. Since the transform involves binomial coefficients, is not
the standard binomial transform of combinatorics, and has a connection to the confluent 
hypergeometric function, we use the name {\it confluent binomial transform}. The reader is welcome to suggest a better name, with appropriate mathematical justification, which we will consider 
 for our future work on this subject. We follow up the discussion of the positive counting 
 formula by describing a construction algorithm for LWPs, which is implemented in Mathematica. 
  In Section \ref{commstar} we show that the product on LWPs, coming from the ring structure in $ \cR / \cI $,  can be expressed as a commutative star product based on a decomposition of the space of all 
   polynomials into LWPs and a transverse space.

 In Section \ref{sec:2moremethods} we give two additional construction methods
 mentioned earlier in the introduction.

Section \ref{sec:future} discusses a number of future research directions 
related to this work.

\section{ Primary fields from differential constraints and Polynomial Rings } \label{RingDescription}

A key result motivating our study is the observation \cite{DRRR17} that primary fields constructed from $n$ copies of a
free scalar $\phi$, along with their derivatives, correspond to polynomials in variables $x_{\mu}^I$ subject to a system of 
linear differential constraints and an $S_n$ invariance condition.  
There are $d$ first order differential constraints coming from the lowest weight condition that $K_{\mu}$ annihilates
a primary field, as well as $n$ Laplacian conditions, coming from the equation of motion.  
We also explain, following 
the statement from \cite{HLMM17}, that these primary fields are in 1-1 correspondence with 
 elements of a polynomial ring,  of which the holomorphic sector  forms a Calabi-Yau ring
 as highlighted in \cite{DRRR17}. The first order constraints take the form of  a
 zero centre of mass momentum condition, when we view these polynomials in $ x_{\mu}^I $ 
 as states in a multi-particle quantum mechanics. They can be solved explicitly. This leads to 
 a formulation of the problem of finding the LWPs as a problem of solving $n$ second order differential 
 constraints acting on  polynomials in $ (n-1)d$  variables, $ X_{ \mu}^A$, where 
 $ 1 \le \mu \le d , 1 \le A \le ( n-1)$. The LWPs, now viewed as polynomials in $\mC [ X_{ \mu}^A ] $, 
 are in 1-1 correspondence with   the elements of a quotient  of $ \mC [ X_{ \mu}^A ] $
 by an ideal generated by quadratic polynomials. The explicit form of these quadratic polynomials 
 is given in terms of Clebsch-Gordan coefficients for the couplings between 
 $V_H \otimes V_H$  and $V_0 \oplus V_H$, where $V_0$ is the trivial representation. 
  
\subsection{ Review : Lowest weight states and primaries from differential equations  }

The scalar field and its derivatives form a vector space $V$, which is an irreducible representation of $ so(d,2)$. This representation is isomorphic to the space of harmonic polynomials in $ x_{\mu}$. The connection between the standard action of the conformal group on the fields and the action of differential operators on the polynomials is explained in \cite{DRRR17}.

The generators of $so(d,2)$ form the algebra
\bea\label{so42} 
&&  [ K_{ \mu } , P_{ \nu} ] = 2 M_{ \mu \nu } - 2 D \delta_{ \mu \nu } \cr 
 && [ D , P_{ \mu} ] = P_{ \mu} \cr 
 && [ D  , K_{ \mu} ] = - K_{ \mu} \cr 
 && [ M_{ \mu \nu } , K_{ \alpha } ] = \delta_{ \nu \alpha } K_{ \mu} - \delta_{ \mu \alpha } K_{ \nu } \cr 
 && [ M_{ \mu \nu } , P_{ \alpha } ] = \delta_{ \nu \alpha } P_{ \mu} - \delta_{ \mu \alpha } P_{ \nu }
 \eea
The algebra  $so(d,2)$ is realised on these polynomials as  \cite{cft4tft2}
\bea\label{diffops}  
&& K_{\mu} = { \partial \over \partial x_{\mu} }\cr 
&& P_{\mu} =  ( x^2\partial_\mu -2x_\mu x\cdot\partial -  ( d -2 )  x_\mu  ) \cr 
&& D = ( x\cdot\partial + {  ( d - 2 ) \over 2 }  ) \cr 
&& M_{\mu \nu} = x_{\mu } \partial_{ \nu} - x_{\nu} \partial_{\mu} \label{pgens}
\eea
Thinking of $x_\mu$ as the co-ordinates of a particle, this is a single particle representation.
The tensor product $V^{ \otimes n } $ can be realized on a many-particle space of functions 
$ \Psi ( x_{\mu}^I ) $, where $ 1 \le I \le n $ labels the particle number. 
We now have generators 
\bea\label{diffopsI} 
&& K_{ \mu}^{ I } = { \partial \over \partial x_{\mu}^I }   \cr 
&& P_{ \mu}^I =  \left(\sum_{\rho=1}^d x^I_{\rho}x^I_\rho {\partial\over \partial x^I_\mu} 
- 2x^I_\mu \sum_{\rho=1}^d x^I_\rho {\partial\over \partial x^I_\rho}
- (d-2) x^I_\mu\right)
\eea
The generators of the diagonal $so(d,2)$ acting as a sum 
of the generators on each tensor factor of $V^{ \otimes n } $ are  
\begin{eqnarray}
   K_\mu = \sum_{ I } K_{ \mu}^{ I } 
\end{eqnarray}
\begin{eqnarray}
P_\mu = \sum_{ I } P_{ \mu}^I 
\end{eqnarray}

Let $ \cH $ be the space of harmonic polynomials in $ x_{\mu}$. 
Polynomials in $ x_{ \mu}^I$ which are harmonic in each of the $ x_{\mu}^I$, i.e. which are annihilated by the $n$ operators 
\bea 
\sum_{\mu=1}^d { \partial^2  \over \partial x^I_{\mu} \partial  x_{\mu}^I  } 
\eea
span the space $ \cH^{ \otimes n } $.

A lowest weight polynomial (LWP) denoted $L ( x_{ \mu}^I ) $ satisfies the equations
\bea\label{eqsLWPs}  
&& \sum_{ I =1 }^n 
 { \partial L   \over \partial x_{\mu}^I  }  = 0 \hbox{ for }   1 \le \mu \le d \cr
&&  \sum_{\mu=1}^d { \partial^2 L  \over \partial x^I_{\mu }  \partial x_{ \mu}^{ I } }   
  =0   \hbox{ for }   1 \le I  \le n  
\eea
The $S_n$ invariant lowest weight polynomials correspond to primary fields. 
We will refer to the first constraint appearing above as the center of mass constraint, for obvious reasons.

In this paper we will focus our attention on LWPs. The projection to $S_n$ invariants 
 is a standard exercise, illustrated in concrete examples in \cite{DRRR17}.

\subsection{ LWPs and the quotient ring $ \cR / \cI$ }\label{sec:quotient-ring} 

The polynomial ring $ \mC [ x_{\mu}^I ] $ is denoted as $ \cR $. 
Consider the ideal $ \cI $ generated by the $ n$ elements $ \sum_{\mu=1}^d 
x^I_{ \mu} x_{ \mu}^I $ along with the $d$ elements $ \sum_{ I } x_{\mu}^I$. 
This is denoted by 
\bea\label{gensI}  
\cI = \langle \sum_{\mu=1}^d x^I_{\mu} x_{ \mu}^I  ,  \sum_{ I } x_{\mu}^I \rangle 
\eea
and consists of elements in $\cR$ of the form 
\bea 
\sum_{ a = 1 }^{ n + d }  h_a g_a 
\eea
where the $g_a $ refer to all the generators in (\ref{gensI}) and $ h_a$ are arbitrary elements of 
the ring $ \cR$. 
Following the statement in \cite{HLMM17}, we will explain that the quotient ring $ \cR / \cI $ is isomorphic as 
a vector space over  $\mC$ to the primaries. $\cR, \cI $ are  vector spaces over $ \mC$ and the quotient 
 ring $ \cR/ \cI$ is also a quotient vector space. 
 Each element is an equivalence class of vectors, related to each other by addition of elements in $\cI$. For each lowest  weight polynomial satisfying (\ref{eqsLWPs})
there is one such equivalence class. It is useful to explain this correspondence. 

Consider the map $ \phi : \cR \rightarrow \cH^{ \otimes n } $ defined by 
\bea 
\phi :  x_{ \mu_1}^{ I_1} x_{\mu_2}^{ I_2} \cdots x_{ \mu_k}^{ I_k} 
\rightarrow P_{ \mu_1}^{ I_1} P_{\mu_2}^{ I_2}  \cdots P_{ \mu_k}^{ I_k} ( 1 ) 
\eea 
$ \sum_{ \mu  }  x_{ \mu}^{I} x_{ \mu}^{ I} $ are in the kernel of this map since, as is easily checked 
using the explicit form of $P_{\mu}^I $ in (\ref{diffopsI}) 
\bea\label{eomI} 
\sum_{\mu=1}^d P_{ \mu}^I P_{ \mu}^I  (1) = 0 
 \eea 
The representation $ \cH^{ \otimes n } $ is, by construction,  a reducible lowest weight representation
of $ so(d,2)$. $1$ is a
lowest weight state for $ SO(d,2)^{ \times n} $  annihilated by $ K_{ \mu}^I$ for all $ I \in \{1, 2, \cdots , n \}$.  
The irrep $\cH_{ n +k , j_1 , j_2 }$ contains  lowest weight states
under the diagonal $ SO(d,2)$  (annihilated by $K_{ \mu} = \sum_I K_{ \mu}^I  $) of dimension 
$\Delta = n\left({d-2\over 2}\right)+k$ and transforming in the rank $k$ traceless symmetric tensor irrep of $so(d)$. 
There will be a multiplicity for each lowest weight state. This is expressed by introducing 
a  vector space of multiplicities $\cM_{ k , j_1 , j_2 }$.
Thus, we can write
\bea 
\cH^{ \otimes n } = \bigoplus_{ k , j_1 , j_2 } \cH_{ n + k , j_1 , j_2 } \otimes \cM_{ k , j_1 , j_2 } 
\eea
For classification of the irreps of $so(d,2)$ and their character formulae, see \cite{Dolan0508}
 and refs therein.  
A lowest weight state with $ \Delta = n +k$ generates a tower of states at higher $ \Delta$ through the action of 
$ P_{ \mu} = \sum_{ I } P_{ \mu}^{ I } $.
These descendants themselves form a subspace that can be characterized as follows
\bea
  {\rm{Descendants}} = { \rm{ Span} }(\cP(\{P_\mu^I\})P_\mu (1))
\eea
where $\cP(\{P_\mu^I\})$ is any polynomial in the $P_\mu^I$. These correspond, under the 
map $ \phi$ to the ideal generated by $x_{\mu} = \sum_I x_{ \mu}^I $. 
The quotient  space $ \cH^{\otimes n}/ {\rm{Descendants}}$ is equivalent, as a vector space, to 
the space of lowest weight states 
\bea 
\cL = \bigoplus_{ k , j_1  , j_2 } \cM_{ k , j_1 , j_2 } 
\eea 
Now consider the homomorphism $ \phi$ as a map from $ \cR $ to $\cH^{\otimes n}/{\rm Descendants}$. 
The kernel of this map is the ideal in $ \cR $ given by the ideal $ \cI $ in (\ref{gensI}). This shows that 
\bea 
\cL =  \cH^{ \otimes n } /  {\rm Descendants }   = \cR / \cI
\eea
Equality here means isomorphism, as graded vector spaces over $ \mC$. 

\subsection{ Representation theory of $V_H$} \label{VHdeveloped}

The $I$ index of $x_\mu^I$, ranging over $ 1 \le I \le n $,
 transforms in the natural representation, $V_{\nat}$ of $S_n$. 
This representation has an orthogonal decomposition into irreducible representations
\bea 
V_{ \nat} = V_0 \oplus V_H 
\eea
$V_0$ is the one-dimensional representation. $V_H$ has dimension $(n-1)$
and corresponds to the Young diagram $[n-1,1]$ with row lengths $n-1,1$. 
The tensor product $V_H \otimes V_H$ can be decomposed into irreducible 
representations as 
\bea 
V_H \otimes V_{H} = V_0 \oplus V_H \oplus V_{ [n-2,2] } \oplus V_{ [n-2,1,1]} 
\eea
The explicit  Clebsch-Gordan coefficients for $V_0$ and $V_H$ will turn out to be useful 
in obtaining a new description of the ring defined earlier in Section \ref{sec:quotient-ring}, 
where  the linear constraints have been solved.

Let us write 
\bea
 V_{\nat}  = \hbox{ Span }  \{ e_1 , e_2  , \cdots , e_n \}
\eea
and introduce the inner product 
\bea\label{natIP}  
\langle e_I , e_J \rangle  = \delta_{ I J } 
\eea
The $S_n$ action on $V_{nat} $ is 
\bea 
D^{ \nat } ( \sigma) e_I = e_{ \sigma^{-1} (I) } 
\eea
and obeys the homomorphism property 
\bea 
D^{ \nat} ( \sigma_1 ) D^{ \nat } ( \sigma_2 )  = D^{ \nat} ( \sigma_1 \sigma_2 ) 
\eea
The inner product (\ref{natIP}) is  invariant under the  $S_n$ action. 
The linear combination
\bea 
e_0 = { 1 \over \sqrt{ n} } \sum_{ I =1}^n e_I 
\eea
is invariant, normalized to $1$  and spans $V_0$.
We can choose a convenient orthonormal basis for $V_H$ as 
\bea\label{YOR}
e_A = { 1 \over \sqrt {  A ( A +1 )}  } 
 ( e_1  + e_2 + \cdots + e_A  - A e_{A+1} )
\eea
for $ A \in \{ 1 , 2 , \cdots  , n-1 \} $. Introducing the notation $ S_{ A I}$ for these 
coefficients we have 
\bea 
e_A = \sum_{ I =1 }^{ n}  S_{A   I } e_I 
\eea
for  $ A \in \{ 1, 2, \cdots , n -1 \}$, and 
\bea 
S_{ A I }  = { 1 \over \sqrt{ A ( A+1)}  }  \left ( - A ~ \delta_{ I , A +1  } + \sum_{ J =1}^{ A } \delta_{ J , I}  \right ) 
\eea 
It is also useful to introduce extend $  A $ to $ A  \in \{ 0 , 1, \cdots , n-1\} $, 
so that 
\bea 
&& e_{  A = 0 } = { 1 \over \sqrt { n } } ( e_1 + e_2 + \cdots + e_n ) \cr
&& S_{ 0  I } = { 1 \over \sqrt n }   \hspace*{2cm} \hbox{for }   ~~ 1 \le I \le n 
\eea
We have the orthonormality relations 
\bea 
\sum_{ I =1}^n S_{  A   I } S_{  B  I } = \delta_{  A  B }  
\eea
This expresses the orthonormality of states $e_{ A} , A \in \{ 1,  \cdots , n-1 \}$ 
within $V_H$, and within $V_0 $ for $A , B = 0 $, as well as the orthogonality of 
all the states in $V_H$ with the invariant state in $V_0$. 
We also have 
\bea 
\sum_{ A=0}^{n-1} S_{ A I} S_{ A   J} = \delta_{ IJ } 
\eea
Given these orthogonality relations, 
the inverse transformation expressing $e_{ I  } $ in terms of  
the $e_A$ are 
\bea\label{invtransf}  
e_{ I } = \sum_{  A =0 }^{ n-1}  S_{  A I  } e_{  A }  
\eea

The following sum will play a crucial role in our subsequent treatment of the ring
defined in Section \ref{sec:quotient-ring}
\bea 
\kappa_{ ABC }  = \sum_{ I =1}^n S_{  C  I }  S_{A I } S_{ B I } 
\eea
Let $D^{ H}_{ C C'} ( \sigma )$ denote 
 the matrix representing the permutation  $\sigma$ in the hook representation $H$. 
Note that  $\kappa_{ABC}$ has the following $S_n$ invariance property. 
\bea 
\kappa_{ABC} &=& \sum_{ I =1}^n \langle H, C | \nat , I \rangle \langle  H,  B | \nat , I \rangle  \langle H, A  | \nat , I \rangle  \cr 
&=& \sum_{ I =1}^n \langle H, C | \nat , \sigma ( I ) \rangle \langle H,  B | \nat , \sigma ( I )  \rangle \langle  H, A  | \nat , \sigma ( I)  \rangle  \cr 
&=& \sum_{ I =1}^n \sum_{ A' , B' , C' =1 }^{ n-1} D^{ H}_{ C C'} ( \sigma )  D^{ H}_{ B  B'} ( \sigma ) D^{ H}_{ A A'} ( \sigma )
 S_{  C'  I }  S_{A' I } S_{ B' I } \cr 
 &=& \sum_{ A' , B' , C' =1 }^{ n-1} D^{ H}_{ C C'} ( \sigma )  D^{ H}_{ B  B'} ( \sigma ) D^{ H}_{ A A'} ( \sigma )
\kappa_{ A'B'C' } 
\eea
This shows that $ \kappa_{ ABC }  $ is a state in $V_H \otimes V_H \otimes V_H$ which is 
invariant under the simultaneous linear transformation  
of the three states by $ S_n$. We know there is, up to normalization, precisely one such 
state, since $V_H$ appears once in the Clebsch-Gordan decomposition of $ V_{H} \otimes V_H $.  Equivalently $V_0$ appears once in $ V_H \otimes V_H \otimes V_H$.  
We conclude that $ \kappa_{ ABC} $ is this Clebsch-Gordan coefficient.

In Appendix \ref{sec:invtVH} we calculate this invariant explicitly to get 
\bea\label{kappaformula} 
&& \kappa_{ A B C } = - ABC \delta_{ A, B , C  } +
  BC \delta_{ B , C } \Theta ( B < A  ) + AB \delta_{ A , B } \Theta ( A < C ) 
 + AC \delta_{ A , C } \Theta ( A < B  )\cr   
&&  - C \Theta ( C < A ) \Theta ( C < B ) - B \Theta ( B < A ) \Theta ( B < C ) - A \Theta ( A < C ) \Theta A < B )  \cr 
&& + Min ( A , B , C )  
\eea
$  \Theta ( B < A )$ is defined to be $1$ if $ B < A $ and $0$ otherwise.  
We also find that 
\bea 
\kappa ( z_A ) & \equiv &   \sum_{A,  B , C =1 }^{ n-1}  \kappa_{ ABC} z_A z_B z_C \cr 
 & =  & \sum_{ A } A ( 1- A^2) z_A^3 + \sum_{ A < B } 3 A ( 1 + A ) z_A^2 z_B  
\eea
and 
\bea 
 \kappa_{ A } ( z ) & = &  \sum_{ B , C } \kappa_{ A B C } z_B z_C \cr 
& = &  A ( 1 - A^2 ) z_A^2 + \sum_{  B : B < A } B ( 1 + B ) z_B^2 
+ \sum_{ B : A < B } 2 A ( 1 + A ) z_B z_A  \cr 
&& 
\eea 

\subsection{ Solving the center of mass constraint and a polynomial ring with quadratic relations } 
\label{sec:quadrels} 

In solving the constraints that determine the LWPs,  a fruitful approach is to solve the
center of mass constraint (COM) and only then consider the remaining constraints in  (\ref{eqsLWPs}).
This approach exploits the $S_n$ structure of the problem. We will use the elements 
of $S_n$ representation theory from Section \ref{VHdeveloped}.

As noted earlier, the $I$ index transforms in the natural representation of $S_n$
which has a decomposition into irreducibles as 
\bea 
V_{ \nat } = V_0 \oplus V_H 
\eea
We will use the  coefficients $S_{ A  I}$ for this decomposition introduced in  Section \ref{VHdeveloped} to define 
\bea 
X^{A}_{\mu}   = \sum_{I=1}^n S_{ A I }  x_{ \mu}^{I} 
\eea
 $X^{0}_{ \mu} $ is invariant under $S_n$. The $X_{ \mu}^{ A}$  for
  $1 \le A \le n-1$ form an orthonormal basis of states in $V_H$.  

The COM condition is satisfied by setting 
\bea\label{lincon} 
X^{0}_{ \mu}   = 0 
\eea
The inverse transformation, following (\ref{invtransf}), is 
\bea 
x_{\mu}^I = \sum_{ A =0 }^{ n-1}   S_{A  I } X_{\mu}^A 
\eea
The quadratic conditions   can be expressed as  
\bea
\sum_{ A , B = 0 }^{ n-1} \sum_{\mu=1}^d  S_{  A  I } S_{  B  I  } X_{\mu}^A X_{ \mu}^B =0 
\eea
The linear COM conditions  (\ref{lincon}) imply the quadratic conditions become 
\bea 
\sum_{ A , B  = 1 }^{ n-1}\sum_{\mu=1}^d  S_{  A  I  } S_{  B  I  } X_{\mu}^A X_{ \mu}^B =0 
\eea
It is useful to express this in the $V_0 \oplus V_H$ basis.
Towards this end, multiply by $S_{C  I } $ and sum over $I$ to find 
\bea 
Q_C\equiv \sum_{I=1}^n \sum_{ A , B  = 1 }^{ n-1} \sum_{\mu=1}^d
S_{ C   I } S_{  A  I  } S_{  B   I }  X_{\mu}^A X_{ \mu}^B =0\label{Qdef} 
\eea
For $ C =0$, we get 
\bea 
Q_0=\sum_{ A }\sum_{\mu=1}^d X_{ \mu}^A X_{ \mu}^A = 0 \label{Qdef0}
\eea
For $C>0$ we have
\bea\label{kappacons} 
\sum_{ \mu =1 }^d  \sum_{ A , B =1  }^{n-1}   \kappa_{ C A B  }   X_{\mu}^A X_{ \mu}^B =0
\eea
Note that, while the $A,B$ indices range over $ \{ 1, \cdots , n-1 \} $, 
the $C$ index ranges over $ \{ 0 , 1, \cdots , n-1 \} $. 
Given the explicit formulae stated in Section \ref{VHdeveloped} and derived 
in Appendix \ref{sec:invtVH}, the quadratic constraints 
can be expressed as 
\bea\label{ExpConsI}  
&&  \hbox { For }  1 \le A \le (n-1)  : \cr 
&& \cr  
&& A ( 1 - A^2 )\sum_{\mu=1}^d X^{ A }_{ \mu} X^{ A }_{ \mu}  + \sum_{ B : B > A   }\sum_{\mu=1}^d 
2 A ( 1+A ) X^{A}_{\mu} X^{ B}_{ \mu} + \sum_{ B : B < A }\sum_{\mu=1}^d 
 B  ( 1+ B  ) X^{B}_{\mu} X^{ B}_{ \mu} = 0 \cr 
&& \hbox{ and } \cr 
&& \cr 
&& \sum_{ A  =1}^{ n-1} \sum_{\mu=1}^d X_{\mu}^{A}  X_{\mu}^{A}=0 
\eea

The upshot is that we have a description of the construction of the primaries as the construction of polynomials in the
hook variables $X_{\mu}^A $  with $ 1 \le A \le n -1 ,  1 \le \mu \le d $, subject to the quadratic constraints (\ref{ExpConsI}). 
In Appendix \ref{conifold} we study the variety defined by these quadratic constraints for some 
low values of $n,d$, and compute the associated Hilbert series
using Sage. It is in complete agreement with our counting of LWPs.

\subsection{$V_0 \oplus V_H$ decomposition of Laplacian constraints. }

The LWPs solve the $n$ Laplacian conditions 
\bea 
\sum_{ \mu } { \partial^2  F \over \partial x_{ \mu}^I  \partial x_{ \mu}^I } = 0 
\eea
These $n$ conditions transform in the natural representation $V_{nat}$ of $S_n$. 
We can again move to the $V_0 \oplus V_{H}$ basis as follows 
\bea 
\Box_{ C }  = \sum_{ I =1 }^n \sum_{\mu=1}^d  S_{C I }  {   \partial^2 \over \partial x^{I}_\mu \partial x^{I}_\mu } 
\eea 
Now expand the $x$-derivatives in terms of $X$-derivatives, and use the fact that 
we are acting on translation invariant functions to drop derivatives with respect to 
$ X^{0}_\mu $. We have 
\bea
\sum_{ I =1 }^n  \sum_{\mu=1}^d S_{ C I  }  {   \partial^2 \over \partial x^{I}_\mu \partial x^{I}_\mu }
=\sum_{ I } \sum_{ A , B =1}^{ n-1} \sum_{\mu=1}^d 
S_{C I}  S_{A I} S_{B I}  {\partial^2 \over \partial X^A_\mu \partial X^B_\mu }  
\eea
Notice that the quantity 
\bea 
\kappa_{ ABC }  = \sum_{ I =1}^n S_{  C  I }  S_{A I } S_{ B I } 
\eea
introduced in the previous section, has appeared above. 

There are $ n-1$ linear combinations of the Laplacians which transform 
as $V_H$, given by 
\bea\label{LapH}  
\Box_C = \sum_{ A , B =1}^{ n-1} \sum_{ \mu =1}^d 
 \kappa_{ A  B  C  } { \partial^2 \over \partial X_{\mu}^{A} \partial X_{\mu} ^{B} } 
\eea
Together with the $S_n$ invariant Laplacian 
\bea\label{Lap0}  
\Box_{ 0 } =  \sum_{\nu =1}^{d} \sum_{  A = 1 }^{ n-1}  { \partial^2   \over \partial x_{ \nu}^{ A } \partial x_{ \nu}^A  } 
\eea
we have $n$ differential operators acting on the functions of $ X_{\mu}^A$.

In summary, we have now arrived at a description of LWPs as polynomial functions in $ X_{\mu}^{A}$ i.e. functions on 
\bea 
( \mR^d)^{ (n-1)} 
\eea
subject to the $n$ Laplacian conditions in (\ref{LapH}) and (\ref{Lap0}).
The LWPs are dual to primary operators in the free CFT.

\section{Counting of lowest weight states in $V^{\otimes n}$}\label{sec:HS}

$V$ is the representation of $so(d,2)$ collecting all the states which correspond, by the operator-state correspondence,
to a single scalar field and its derivatives. Above we have established that the lowest weight states in $V^{\otimes n}$ 
form a polynomial ring. We will develop this description further in this section by counting these lowest weight states. 
Specifically, we give a formula for the generating  function of the number of lowest weight states in $V^{\otimes n}$, 
at weight $\Delta=n\left({d-2\over 2}\right)+k$. The $S_n$ invariant states among these lowest weight states are the primaries.

In $d$ dimensions, the character of the free scalar field $so(d,2)$  irrep is 
\bea 
\chi_V (s)  = tr s^{ \Delta }  = s^{(d -2)/2}  { ( 1  - s^2 ) \over ( 1 - s)^{ d } }
\eea
The character for $V^{\otimes n}$, a reducible representation, is then 
\bea 
\chi_{V^{\otimes n}}(s)=(\chi_V (s) )^n =s^{ n(d-2)/2}  { ( 1  - s^2 )^n \over ( 1 - s)^{ nd } }
\eea
The trace of $ s^{ \Delta }$ over states  obtained by acting with momenta  on a lowest weight state (annihilated by $ K_{\mu}$) with  
$\Delta = n(d -2)/2+k$ is 
\bea\label{trdes}  
{s^{ n(d -2)/2 + k } \over ( 1 - s )^{ d } }  
\eea
Let  the multiplicities of these lowest weight states  in $ V^{ \otimes n } $ be $\cN_{ k }$.
To determine the multiplicities $\cN_{k}$ we expand $\chi_{V^{\otimes n}}(s)$ 
in terms of the traces in (\ref{trdes}) as follows 
\bea 
\chi_{V^{\otimes n}}(s)=s^{ n(d -2)/2}  { ( 1  - s^2 )^n  \over ( 1 - s)^{ n d  } } =
 \sum_{ k=0}^{ \infty } \cN_k { s^{ n(d -2)/2 + k } \over ( 1 - s )^{ d } }  
\eea
Hence the generating function for the multiplicities of lowest weight states is 
\bea\label{charprediction}  
\sum_{ k=0}^{ \infty } \cN_k  s^{ k } && = {  ( 1 - s^2 )^n \over ( 1 - s)^{ d ( n -1) } } \cr 
&& = { 1 - n s^2 +  { n ( n-1) \over 2 } s^4 + \cdots  \over ( 1 - s)^{ d ( n -1) } } \cr 
&& =  { 1\over ( 1 - s)^{ d ( n -1) } } \sum_{ k=0  }^n   (-1)^k { n \choose k } s^{ 2k} 
\eea
In this result we can already recognize elements of our discussion from section \ref{RingDescription} appearing.
Indeed, the denominator of the Hilbert series given above shows that there are $d(n-1)$ generators in the ring.
These are the $X_{\mu}^{A}$. The numerator implies that at quadratic order, we have $n$ relations.
These are the constraints (\ref{Qdef0}) and (\ref{kappacons}).
Note that $ { n \choose k }  $ is the dimension of the $k$-fold anti-symmetric product of 
$V_Q$, the $n$ dimensional space spanned by the quadratic constraints $Q_A$. An important point 
for the discussion in the next section is that 
\bea 
{ 1\over ( 1 - s)^{ d ( n -1) } }  { n \choose k } s^{ 2k}
\eea
is the trace of $s^{ \Delta }$ over $ \cR \otimes \Lambda^k ( V_Q ) $. 
Finally, it is worth noting that the counting function in  the first line of 
(\ref{charprediction}) is palindromic.

 \section{  The ring of lowest weights in $ V^{ \otimes n } $ }\label{sec:exactsequences}

In the previous section we have obtained the counting function for the lowest weights in $V^{\otimes n}$. 
These lowest weights form a polynomial ring. The counting function for the ring 
is a rational function. The ring is a quotient of the polynomial ring, by an ideal. 
The ideal is generated by $n$ quadratic expressions. 

The structure of the counting function can be explained using the theory of Hilbert series, in terms of the relations between the generators of the ideal, relations between these relations and so on. This notion of generators, relations and relations between relations is made precise in the theory of Hilbert series in terms of exact sequences of modules over the polynomial ring. References we found useful include \cite{Cox,Eisenbud,Grigorescu}.

In this section we will describe the relevant exact sequence and show that it matches the counting function derived from 
$so(d,2)$ representation theory.  
We then explain how the exact sequence of modules of the polynomial ring $ \cR$
leads to exact sequences of vector spaces over the base field $ \mC$. 
These exact sequences are used to derive a refined counting formula
for $so(d) \times S_n$ irreps among the lowest weights. 
The $so(d) $ scalar lowest weights are of interest in effective field theory \cite{HLMM17}. 

\subsection{ Exact sequence of modules } \label{modulesequence}

We will consider the following exact sequence of modules over $\cR = \mC [ X_{\mu}^{A}]$ 
\bea\label{seqmodules}  
&& 0 \xrightarrow{f_0}  \cR \otimes \Lambda^{ n } ( V_Q )   \xrightarrow{f_n }   \cdots  \rightarrow   \cR \otimes \Lambda^2 ( V_Q ) \xrightarrow{f_2}  \cR \otimes V_Q    \xrightarrow{f_1}   \cR  \xrightarrow{f_{\cR}}   \cL  \xrightarrow{f_{\cL }} 0 \cr 
&& 
\eea 
The tensor products are defined over the base field $ \mC$. Elements of  $\cR\otimes V_Q$ are given by 
\bea 
\sum_{ A=0}^{ n-1} h_A \otimes Q_A
\eea
where $ h_A \in \cR $. The map $f_1$ acts as 
\bea 
f_1 : \sum_{ A=0}^{ n-1} h_A \otimes Q_A \rightarrow \sum_{ A=0}^{ n-1} h_A Q_A
\eea 
Its image is the ideal generated by $Q_A$. This ideal $\cI (d,n)$,  consists of elements in the ring $\cR =\mC [X_{\mu}^A]$ of the form 
\bea 
\sum_{A=0}^{n-1} h_A Q_A 
\eea
where $h_A $ are general elements in $ \cR (d,n)  $.

 $ \cR \otimes V_Q $  is a module for $ \cR$. 
An element $h \in \cR $ acts as 
\bea 
\sum_{ A=0}^{ n-1} h_A \otimes Q_A \rightarrow \sum_{ A=0}^{ n-1} h h_A \otimes Q_A 
\eea
Since $ \cL $ is the quotient $ \cR / \cI$ and the map $ f_{ \cL } $ takes all the elements of $ \cL$ to $0$, the image of $ f_{\cR} $ is the kernel of $ f_{ \cL} $. 
Thus the sequence is exact at $\cR $ and $\cL$.  

The elements of $ \cR \otimes \Lambda^I ( V_Q ) $ are 
\bea
\epsilon^{ A_1, \cdots , A_I, A_{ I+1}, \cdots  , A_n } h_{ A_1 A_2 \cdots A_I } \otimes Q_{A_1} \otimes Q_{A_2}  \cdots \otimes Q_{ A_I} 
\eea
(the repeated indices $A_{I+1} , \cdots , A_{n} $ are summed). 
Under the map $ f_I$, they go to 
\bea 
\epsilon^{ A_1, \cdots , A_I, A_{ I+1}, \cdots  , A_n } h_{ A_1 A_2 \cdots A_I }  Q_{A_1} \otimes Q_{A_2}  \cdots \otimes Q_{ A_I} 
\eea 
Under the composite map $ f_{ I } \circ f_{ I-1}$, we have
\bea
&& f_{ I } \circ f_{ I-1} : 
\epsilon^{ A_1, \cdots , A_I, A_{ I+1}, \cdots  , A_n } h_{ A_1 A_2 \cdots A_I } \otimes Q_{A_1} \otimes Q_{A_2}  \cdots \otimes Q_{ A_I} 
\rightarrow  \cr 
&& \epsilon^{ A_1, \cdots , A_I, A_{ I+1}, \cdots  , A_n } h_{ A_1 A_2 \cdots A_I }  Q_{A_1}  Q_{A_2} \otimes Q_{ A_3}   \cdots \otimes Q_{ A_I}  =  0 
\eea
This shows that 
\bea
Im ( f_{ I } ) \subseteq Ker ( f_{ I-1} ) \label{ImKer}
\eea
Thus we have established that the image of $f_I$ is in the kernel of $f_{I-1}$.
For exactness we need to show that the image of $f_I$ is equal to the kernel of $f_{I-1}$.
This will follow, by additionally proving that $ Ker ( f_{ I-1} )\subseteq  Im ( f_{ I } )$.

To proceed further, motivated by the analysis in Chapter 4 of \cite{WuGang} (where 
exactness is proved for a sequence involving $ \Sym^{k} ( V ) \otimes \Lambda^k ( V ) $, 
in the context of proving that $\Sym(V)$ and $ \Lambda(V)$ are Koszul algebras), 
 we introduce two operators.
The first operator, $d$, is a re-expression of the maps $f_I$ introduced above.
We define
\bea
   d=\sum_{A=0}^{n-1} Q_A \otimes \iota_A
\eea
where $Q_A$ acts by multiplying any polynomial $f$ in $\cR$ by $\sum_{\mu=1}^d\kappa_{ABC}X^B_\mu X^C_\mu$ and
$\iota_A$ is interior multiplication with $Q_A$
\bea
\iota_A(Q_{B_1}\wedge\cdots\wedge Q_{B_i})&=&
\sum_{k=1}^i (-1)^{k-i} Q_{B_1}\wedge\cdots\wedge \hat{Q}_{B_k}\wedge \cdots \wedge Q_{B_i}
\,\,\,{\rm if}\,\,\, A=B_k\cr\cr
&=& 0 \qquad A\notin\{B_1,B_2,\cdots,B_i\}
\eea
It is simple to demonstrate, using this definition, that $d^2=0$, which is equivalent to the discussion above
that lead to the conclusion (\ref{ImKer}).
To motivate the second operator we need, we employ a decomposition of polynomials that will be derived in
Section \ref{natinner}: the space of degree $d$ polynomials can be decomposed as
\bea 
   p^{(d)} = p_h^{(d)} + Q_A p_{h,A}^{ (d-2) } + Q_AQ_B p_{h,AB}^{(d-4)} + \cdots \label{decomp}
\eea
with the coefficients $p_h^{(d)}$, $p_{h,A}^{(d-2)}$, $p_{h,AB}^{(d-4)}$ all annihilated by ($ C=0,1,...,n-1$). 
\bea
   \Box_{ C} = \sum_{ A , B =1}^{ n-1} \sum_{ \mu =1}^d
 \kappa_{ CAB} { \partial^2 \over \partial X_{\mu}^{A} \partial X_{\mu} ^{B} } 
\eea
The repeated index $ A$ in the second term is summed, as are the $A,B$ in the third term, etc.  
The decomposition (\ref{decomp}) is unique in the sense that the coefficients $p_h^{(d)}$, $p_{h,A}^{(d-2)}$, 
$p_{h,AB}^{(d-4)}$,..., in the expansion are unique.
The second operator we use is
\bea
   \alpha = \sum_{A=0}^{n-1} d_A \otimes\psi_A
\eea
where
\bea
   \psi_A(g)=g\wedge Q_A
\eea
and the action of $d_A$ is defined using the expansion (\ref{decomp}): $d_A$ simply removes a $Q_A$ from each
term and it annihilates $p_h^{(d)}$:

\bea
d_A \left(p_h^{(d)} + Q_B p_{h,B}^{ (d-2) } + Q_BQ_C p_{h,BC}^{(d-4)} + \cdots\right)
=p_{h,A}^{(d-2)}+2 Q_B p_{h,AB}^{(d-4)}+\cdots\cr
\eea
Again, from these definitions it is easy to see that $\alpha^2=0$: applying $d_A$ twice produces a symmetric two 
index tensor $p^{q}_{h,ABC\cdots D}Q_C\cdots Q_D$ which vanishes when summed against 
$g\wedge Q_A\wedge Q_B$, which is antisymmetric in $A,B$.

We will now argue that $\alpha\circ d+d\circ \alpha$ is not degenerate, that is, it has an inverse.
In fact, we will show that acting on a monomial of degree $t$ in the $Q$s, it is proportional to the identity
\bea
    \alpha\circ d+d\circ \alpha =  1+t
\eea
In this case, acting on any element 
$k^{(q)}_{h,AB\cdots E}Q_AQ_B\cdots Q_E\otimes Q_{A_1}\wedge \cdots\wedge Q_{A_i}$ in the kernel of $d$, we have
\bea
(\alpha\circ d+d\circ \alpha)k^{(q)}_{h,AB\cdots E}Q_AQ_B\cdots Q_E\otimes Q_{A_1}\wedge \cdots\wedge Q_{A_i}\cr
=d\circ \alpha (k^{(q)}_{h,AB\cdots E}Q_AQ_B\cdots Q_E\otimes Q_{A_1}\wedge \cdots\wedge Q_{A_i})\cr
=d \left(\alpha (k^{(q)}_{h,AB\cdots E}Q_AQ_B\cdots Q_E\otimes Q_{A_1}\wedge \cdots\wedge Q_{A_i})\right)\cr
=(1+t)k^{(q)}_{h,AB\cdots E}Q_AQ_B\cdots Q_E\otimes Q_{A_1}\wedge \cdots\wedge Q_{A_i}
\eea
so that
\bea 
Ker ( f_{ I-1} ) \subseteq Im ( f_{ I } ) 
\eea
and hence, together with (\ref{ImKer}) we have exactness 
\bea 
Ker ( f_{ I-1} ) =  Im ( f_{ I } ) 
\eea

Let us now complete the argument by showing that $\alpha\circ d+d\circ \alpha$ is indeed not degenerate.
For $d\circ\alpha$ we find (assume that $p^{(q)}_{h,AB\cdots E}$ has $t$ indices $AB\cdots E$)
\bea
&&d\circ\alpha (p^{(q)}_{h,AB\cdots E}Q_AQ_B\cdots Q_E\otimes Q_{A_1}\wedge \cdots\wedge Q_{A_i})\cr\cr
&=& d\left(\sum_{S\notin \{ A_1,...,A_i\}}
t p^{(q)}_{h,SB\cdots E}Q_B\cdots Q_E\otimes
Q_{A_1}\wedge \cdots\wedge Q_{A_i}\wedge Q_S\right)\cr\cr
&=&\sum_{S\notin \{ A_1,...,A_i\}}
t Q_S p^{(q)}_{h,SB\cdots E}Q_B\cdots Q_E\otimes
Q_{A_1}\wedge \cdots\wedge Q_{A_i}\cr\cr
&+&\sum_{S\notin \{ A_1,...,A_i\}}\sum_{k=1}^i (-1)^{i-k+1}tQ_{A_k}
p^{(q)}_{h,SB\cdots E}Q_B\cdots Q_E\otimes
Q_{A_1}\wedge \cdots\hat{Q}_{A_k}\cdots\wedge Q_{A_i}\wedge Q_S\cr
&&\label{da2}
\eea
For $\alpha\circ d$ we have
\bea
&&\alpha\circ d (p^{(q)}_{h,AB\cdots E}Q_AQ_B\cdots Q_E\otimes Q_{A_1}\wedge \cdots\wedge Q_{A_i})\cr
&=& \alpha \left(\sum_{k=1}^i (-1)^{i-k} p^{(q)}_{h,AB\cdots E}Q_AQ_B\cdots Q_E Q_{A_k}
\otimes
Q_{A_1}\wedge \cdots\hat{Q}_{A_k}\cdots\wedge Q_{A_i}\right)\cr
&=&\sum_{k=1}^i \sum_{S\notin \{ A_1,...,\hat{A}_k,...,A_i\}} (-1)^{i-k}
d_S(p^{(q)}_{h,AB\cdots E}Q_AQ_B\cdots Q_E Q_{A_k})\otimes
Q_{A_1}\wedge\cdots\hat{Q}_{A_k}\cdots\wedge Q_{A_i}\wedge Q_S\cr\cr
&=&\sum_{k=1}^i\sum_{S\notin \{ A_1,...,\hat{A}_k,...,A_i\}} (-1)^{i-k}
t p^{(q)}_{h,SB\cdots E}Q_B\cdots Q_E Q_{A_k}\otimes
Q_{A_1}\wedge\cdots\hat{Q}_{A_k}\cdots\wedge Q_{A_i}\wedge Q_S\cr\cr
&+&p^{(q)}_{h,AB\cdots E}Q_AQ_B\cdots Q_E\otimes
Q_{A_1}\wedge\cdots\wedge Q_{A_i}\label{ad2}
\eea
The second term in the answer (\ref{da2}) cancels against the first term in (\ref{ad2}) to leave
\bea
\sum_{k=1}^i 
t p^{(q)}_{h,A_kB\cdots E}Q_B\cdots Q_E Q_{A_k}\otimes
Q_{A_1}\wedge\cdots\wedge Q_{A_i}\cr
\eea
Thus, we find
\bea
&&(d\circ\alpha +\alpha\circ d)(p^{(q)}_{h,AB\cdots E}Q_AQ_B\cdots Q_E\otimes Q_{A_1}\wedge 
\cdots\wedge Q_{A_i})\cr\cr
&=& \sum_{S\notin \{ A_1,...,A_i\}}
t p^{(q)}_{h,SB\cdots E}Q_S Q_B\cdots Q_E\otimes
Q_{A_1}\wedge \cdots\wedge Q_{A_i}\cr\cr
&+&\sum_{k=1}^i t p^{(q)}_{h,A_kB\cdots E}Q_B\cdots Q_E Q_{A_k}\otimes
Q_{A_1}\wedge\cdots\wedge Q_{A_i}\cr\cr
&+&
p^{(q)}_{h,AB\cdots E}Q_AQ_B\cdots Q_E\otimes
Q_{A_1}\wedge\cdots\wedge Q_{A_i}\cr
&=&(1+t)
p^{(q)}_{h,AB\cdots E}Q_AQ_B\cdots Q_E\otimes
Q_{A_1}\wedge\cdots\wedge Q_{A_i}
\eea
which is the result we wanted.

Finally, note that the above long exact sequence can be thought of as consisting of several short exact sequences. 
One is the standard short exact sequence for quotients
\bea 
0 \rightarrow \cI \rightarrow \cR  \rightarrow \cR/ \cI \rightarrow 0 
\eea
The next is 
\bea 
 Syz ( \cI ) \rightarrow \cR \otimes V_Q \rightarrow \cI 
\eea
Here a basis for $V_Q$ gives the generators of $ \cI$. $  Syz ( \cI )  $ is the syzygy module for $ \cI$.  It is generated by $ \Lambda^2 ( V_Q)$ so that its generators 
and relations are expressed in a sequence 
\bea 
  Syz ( Syz ( \cI ) ) \rightarrow \cR \otimes \Lambda^2 ( V_Q) \rightarrow Syz ( \cI ) 
\eea

\subsection{ Exact sequence of vector spaces over $ \mC $ } 

We can consider the vector space formed by polynomials of a fixed degree.
Since the modules of the last section are defined over the ring, a single exact sequence of modules implies, upon specializing to fixed degree, an exact sequence for each of these vector spaces. 
The polynomials at fixed $n$ and fixed degree $k$ are polynomials dual to primaries constructed using $n$ fields and $k$ derivatives.
There is therefore an interesting CFT motivation to consider the exact sequences between the vector spaces formed by polynomials of a fixed degree, which is the goal of this section.
The $X_{\mu}^A$ transform as $ V_d \otimes V_H$ of $so(d) \times S_n$, where $V_d$ is 
the $d$-dimensional vector of $so(d)$ and $V_H$ is the $(n-1)$ dimensional hook representation of $S_n$. 
For convenience, we define $ V_{dH}= V_d \otimes V_H$. 
Polynomials of degree $k$ in $ X_{ \mu}^A$ form a vector space over $ \mC$ isomorphic to the 
space of rank $k$ symmetric tensors, denoted $\Sym^k ( V_{dH} ) $. 

At $k=2$ we have
\bea 
0 \rightarrow V_Q \rightarrow \Sym^2 ( V_{ dH } ) \rightarrow \cL( 2 , d, n ) \rightarrow 0 \label{deg2es}
\eea
The space of LWPs at $k=2$, denoted $ \cL (2,d,n)$, is obtained by setting the $Q_A$'s to zero. 
The space $\Sym^2 (V_{ dH })$ is the space of degree two polynomials in the hook variables $X^A_\mu$.
The $Q_A$'s form a subspace of $\Sym^2 (V_{dH}) $, so we have a map from $V_Q$ to  $\Sym^2(V_{dH})$. 
The definition of $\cL (2,d,n)$ as the quotient space ensures that the image of the first map is exactly the kernel of the second map, so that the sequence above is exact. 
Denoting the dimension of $\Sym^k (V_{dH})$ by $S (k,d,n)$, we know that
\bea
S(k,d,n)={ ( ( n-1) d + k -1 ) ! \over k! ( ( n-1) d -1 ) !  } \equiv S ( k , D = d ( n-1) )\label{DimSim}
\eea
The second equality emphasizes the fact that this depends only on $D=d(n-1)$.
The exact sequence (\ref{deg2es}) implies the following formula for the dimension of the space of LWPs
\bea 
L (2, d, n) = S (2, d , n ) - \Dim ( V_Q) 
\eea 
It is simple to check this independently by comparing to the coefficient of $s^2$ in the expansion of (\ref{charprediction}).

Next, consider degree $k=4$. The relevant exact sequence is
\bea 
0 \rightarrow  \Lambda^2 ( V_Q)  \rightarrow  \Sym^2 ( V_{ dH} ) \otimes V_Q  \rightarrow \Sym^4 ( V_{ dH } ) \rightarrow \cL( 4 , d, n ) \rightarrow 0 \label{deg4es}
\eea
as we explain below. Start by introducing the map $f$ defined by
\bea 
f :  \Sym^2 ( V_{dH} ) \otimes  V_Q  \rightarrow \Sym^4 ( V_{dH} ) 
\eea
Concretely, we have
\bea 
f : Q_A \otimes X_{ \mu_1}^{ a_1} X_{ \mu_2}^{ a_2 }  \rightarrow Q_{ A}  X_{ \mu_1}^{ a_1} X_{ \mu_1}^{ a_2 } = \sum_{\mu=1}^d \kappa_{A B C }  X_{ \mu}^{ B } X_{ \mu}^{ C }X_{ \mu_1}^{ a_1} X_{ \mu_2}^{ a_2 }
\eea
Thus, under this map we find
\bea
f : Q_A \otimes Q_B - Q_B \otimes Q_A  \rightarrow 0 
\eea
Thus, the kernel of the map is $\Lambda^2 (V_Q)$. 
When we have  a 4-term exact sequence as above the far right vector space is the cokernel, i.e. 
\bea 
\cL ( 4 , d ,n   ) = \Sym^4 ( V_{dH} ) / Im ( f ) 
\eea
This is indeed the definition of $ \cL (4,d,n)$: It is the quotient space of the degree $4$ polynomials in $ X_\mu^A$ obtained by setting to zero anything of the form $QXX$. 
The exact sequence (\ref{deg4es}) implies the following relation 
\bea 
L ( 4,d , n ) = S ( 4 , d, n) - S ( 2,d,n) \Dim ( V_Q  )  + \Dim \Lambda^2 ( V_Q) 
\eea
which agrees with the coefficient of $s^4$ in the expansion of (\ref{charprediction}).

These exact sequences generalize to any $k$.  We have 
\bea\label{ESevenk} 
&&  0 \rightarrow \Sym^{ k - 2L } ( V_{ dH} ) \otimes \Lambda^{ L   } ( V_Q) \rightarrow \cdots \rightarrow 
\Sym^{ k - 2I } ( V_{ dH}  ) \otimes  \Lambda^{  I   } ( V_Q) \rightarrow \cdots \cr 
&&  \rightarrow 
\Sym^{ k - 2 } ( V_{ dH}  ) \otimes   V_Q \rightarrow 
\Sym^{ k } ( V_{ dH } ) \rightarrow \cL ( k , d , n ) \rightarrow 0 \cr 
&& 
\eea 
where $ L = min ( \lfloor { k \over 2 } \rfloor  , n ) $. 
If $ k $ is even and $ k/2 \le n $, then the second term in the sequence is 
 $ \Lambda^{ k/2} ( V_Q) $. If $k/2 \ge n $, it is $ \Sym^{k-2n}(V_{dH}) \otimes \Lambda^{ n } ( V_Q) $. If $ k $ is odd and $ (k-1)/2 \le n $, then the first non-trivial term 
 is $ V_{ dH } \otimes \Lambda^{ { k -1 \over 2 }  } ( V_Q)$. If $ n \le (k-1)/2$, 
 then it is $ \Sym^{k-2n}(V_{dH}) \otimes \Lambda^{ n } ( V_Q) $.

One basic building block that the above  sequences are constructed from is the following
\bea
\cdots\rightarrow \Sym^{k-2I}(V_{dH})\otimes\Lambda^{I}(V_Q)
\rightarrow \Sym^{k+2-2I}(V_{dH})\otimes\Lambda^{I-1}(V_Q)\cr
\rightarrow \Sym^{k+4-2I}(V_{dH})\otimes\Lambda^{I-2}(V_Q)
\cdots
\eea
A simple generalization of the discussion above gives the maps required for this basic building block.
First note that the space $\Lambda^{I-1}(V_Q)$ is spanned by 
\bea
\epsilon^{A_1\cdots A_{I-1}A_I\cdots A_L}Q_{A_1}\otimes Q_{A_2}\otimes\cdots\otimes Q_{A_{I-1}}
\eea
with $ L = \min ( \lfloor { k \over 2 } \rfloor  , n ) $ 
as above. 
Define the map $f$ which maps
\bea
f:\Sym^{k+2-2I}(V_{dH})\otimes\Lambda^{I-1}(V_Q)
\rightarrow \Sym^{k+4-2I}(V_{dH})\otimes\Lambda^{I-2}(V_Q)
\eea
as follows
\bea
&&f(X^{a_1}_{\mu_1}\cdots X^{a_{k+2-2I}}_{\mu_{k+2-2I}}
\otimes \epsilon^{A_1\cdots A_{I-1}A_I\cdots A_L}Q_{A_1}\otimes Q_{A_2}\otimes\cdots\otimes Q_{A_{I-1}})\cr
&&=\sum_{\mu=1}^d X^{a_1}_{\mu_1}\cdots X^{a_{k+2-2I}}_{\mu_{k+2-2I}}\kappa_{A_{I-1}BC}X^B_\mu X^C_\mu
\otimes \epsilon^{A_1\cdots A_{I-1}A_I\cdots A_L}Q_{A_1}\otimes Q_{A_2}\otimes\cdots\otimes Q_{A_{I-2}}
\nonumber
\eea
It is clear that the image of the map
\bea
g:\Sym^{k-2I}(V_{dH})\otimes\Lambda^{I}(V_Q)
\rightarrow \Sym^{k+2-2I}(V_{dH})\otimes\Lambda^{I-1}(V_Q)
\eea
is in the kernel of $f$.
Indeed, the image of $g$ is spanned by
\bea
\sum_{\mu=1}^d X^{a_1}_{\mu_1}\cdots X^{a_{k-2I}}_{\mu_{k-2I}}\kappa_{A_{I}BC}X^B_\mu X^C_\mu
\otimes \epsilon^{A_1\cdots A_{I-1}A_I\cdots A_L}Q_{A_1}\otimes Q_{A_2}\otimes\cdots\otimes Q_{A_{I-1}}\cr
\eea
Under $f$ this maps to zero
\bea
&&f(\sum_{\mu=1}^d X^{a_1}_{\mu_1}\cdots X^{a_{k-2I}}_{\mu_{k-2I}}\kappa_{A_{I}BC}X^B_\mu X^C_\mu
\otimes \epsilon^{A_1\cdots A_{I-1}A_I\cdots A_L}Q_{A_1}\otimes Q_{A_2}\otimes\cdots\otimes Q_{A_{I-1}})\cr\cr
&&=\sum_{\mu,\nu=1}^d X^{a_1}_{\mu_1}\cdots X^{a_{k-2I}}_{\mu_{k-2I}}\kappa_{A_{I}BC}X^B_\mu X^C_\mu
\kappa_{A_{I-1}FG}X^F_\nu X^G_\nu \otimes\cr\cr
&&\qquad\qquad\qquad  \epsilon^{A_1\cdots A_{I-1}A_I\cdots A_L}Q_{A_1}\otimes Q_{A_2}\otimes\cdots\otimes Q_{A_{I-2}}=0
\eea
where the last equality follows because
\bea
\sum_{\mu,\nu=1}^d \kappa_{A_{I}BC}X^B_\mu X^C_\mu\kappa_{A_{I-1}FG}X^F_\nu X^G_\nu
\eea
is symmetric under swapping $A_I$ and $A_{I-1}$, and it is contracted with $\epsilon^{A_1\cdots A_L}$.
To complete the discussion, consider
\bea
\cdots\rightarrow \Sym^{k-2}(V_{dH})\otimes V_Q\rightarrow \Sym^{k}(V_{dH})\rightarrow{\cal L} (k,d,n)\rightarrow 0
\eea
In terms of the map $h$ which maps
\bea
h:\Sym^{k-2}(V_{dH})\otimes V_Q\rightarrow \Sym^{k}(V_{dH})
\eea
we have ${\cal L} (k,d,n)=\Sym^{k}(V_{dH})/Im(h)$, which is true since  ${\cal L}(k,d,n)$ is the quotient space of the degree $k$ polynomials in the $X_{\mu}^A$ obtained by setting anything of the form $QX\cdots X$ to zero. 

The argument above shows that the image is in the kernel. 
To establish exactness, we need to show that the kernel is equal to the image. 
This can be done exactly as we did it in Section \ref{modulesequence}.
We again introduce $d$ (again motivated by the mappings we just discussed) and $\alpha$, defined precisely
as we did above.
The generalization of the argument is then obvious and we will not repeat it here.

The exact sequences we have presented in this section imply that
\bea\label{altsumL1} 
L(k,d,n) = \sum_{I=0}^{min ( \lfloor{k\over 2}\rfloor , n ) }(-1)^I \Dim(\Sym^{k-2I}(V_{dH})) \Dim (\Lambda^I (V_Q))
\eea
This formula will be used in the next section to refine the counting of LWPs, by keeping track of the $so(d)\times S_n$ irreps of the LWPs. To understand why this refined counting is possible, note that the maps involved in the exact sequences given in this section, all commute with $so(d)\times S_n$.
The maps involved in the vector space exact sequences involve replacing $Q_A$ by its explicit form $\kappa_{ABC}X^B_\mu X^C_\mu$. Since the spacetime indices are fully contracted, the maps replaces an $so(d)$ scalar with an $so(d)$ scalar.
Similarly, since $\kappa_{ABC}$ is an invariant tensor, the map is from the hook to the hook irrep.
Since this refinement holds for all of the vector space exact sequences, it should hold for the module exact sequences too. This is indeed clear from the expressions $ d = \sum_{ A } Q_A \otimes \iota_A $ 
used in Section \ref{modulesequence}.

\section{ Refined counting formulae }\label{sec:refinedcounting} 

We have managed to count the number of LWPs of fixed degree, or equivalently, 
lowest weight states in $V^{ \otimes n }$. 
There are good reasons to refine this counting using the $so(d) \times S_n$ symmetry present in the problem. Primaries in the free field theory are $S_n$ invariants. They 
 are labeled by their dimension and $so(d)$ representation property.
  In addition, $so(d)$ scalars are relevant
for identifying possible terms in the Lagrangian of effective field theory \cite{HLMM17}. 
This refined counting will ultimately lead to a construction algorithm for the LWPs.
In this section we will carry out this refined counting, using the formula (\ref{altsumL1}) which follows from the exact sequences developed in the last section.
 
A useful starting point for the refined counting is (\ref{altsumL1}) which we re-write slightly 
here, by substituting $ V_Q \rightarrow V_{\nat } $ 
\bea\label{altsumL} 
 L ( k , d , n ) = \sum_{ I =0 }^{\min ( \lfloor{k\over 2}\rfloor , n ) } ( -1)^I \Dim ( \Sym^{ k - 2I }  ( V_{ dH } ) ) \Dim ( \Lambda^I  ( V_{ \nat }^{ ( S_n ) }  ) )
\eea
We can start by writing 
\bea
\Sym^{k-2I}(V_{dH})=\bigoplus_{\Lambda_1,\Lambda_{3,1}\vdash n}
V_{\Lambda_1}^{(so(d))}\otimes V_{\Lambda_{3,1}}^{(S_n)}
\otimes V_{\Lambda_1,\Lambda_{3,1}}
\eea
The RHS is the decomposition of $\Sym^{k-2I}(V_{dH}) $ in terms of 
irreducible representations of $so (d) \times S_n$, labeled by $ ( \Lambda_1  , \Lambda_{3,1} )$. 
$\Lambda_{3,1}$ is a partition of $n$. 
A basis in terms of irreps will include a multiplicity label 
for the pair $ ( \Lambda_1  , \Lambda_{3,1} )$, this multiplicity space is denoted 
by $ V_{\Lambda_1,\Lambda_{3,1}} $.

We will denote the dimensions of these multiplicity spaces  by\\ 
 $\Mult((\Sym^{k - 2I}(V_{dH});\Lambda_1^{so(d)}\otimes\Lambda_{3,1}^{ (S_n)})$.
We can also decompose the antisymmetric (wedge) product of $n$ copies of the natural representation $\Lambda^I (V_{ nat }^{(S_n)})$ into irreps of $S_n$.
The number of times a given $S_n$ irrep $\Lambda_{3,2}$ appears will be denoted by $\Mult(\Lambda^I(V_{\nat}^{(S_n)}); \Lambda_{3,2}^{(S_n)})$.
We can now write the $ so(d) \times S_n$ refined version of (\ref{altsumL}) as
\bea\label{sodsnrefined}  
&&  L ( \Lambda_1^{ so(d)}  , \Lambda_3^{ (S_n)} ; k , d , n ) \cr 
 && 
 = \sum_{ \Lambda_{3 , 1 }, \Lambda_{ 3,2} \vdash n  }  \sum_{ I = 0 }^{ min ( \lfloor{k\over 2}\rfloor , n )  }  (-1)^I \Mult ( ( \Sym^{k - 2I }  ( V_{ dH } ) ;  \Lambda_1^{ so(d)} \otimes  \Lambda_{3,1}^{ (S_n)} ) \cr 
 && \hspace*{2cm}   \Mult ( \Lambda^I ( V_{ \nat }^{ ( S_n ) }  ) ; \Lambda_{3,2}^{ ( S_n ) } )
 C ( \Lambda_{ 3,1} , \Lambda_{ 3,2} ; \Lambda_3 )   \cr 
 && 
 \eea     
$ C ( \Lambda_{ 3,1} ,   \Lambda_{ 3,2} ; \Lambda_3 )  $ is the Kronecker multiplicity for 
$ \Lambda_{ 3,1} \otimes \Lambda_{ 3,2} \rightarrow \Lambda_3 $.     
The LHS is the multiplicity of irreps  $ \Lambda_1 , \Lambda_3 $  in the space of lowest weight states 
of dimension $ L ( k, d, n ) $. Consequently, we have
\bea 
L ( k , d, n ) = \sum_{ \Lambda_1 , \Lambda_3 } \Dim_{so(d)}  ( \Lambda_1 )  \Dim_{S_n } ( \Lambda_3 ) 
  L ( \Lambda_1^{so(d)}  , \Lambda_3^{(S_n)} ; k , d , n )
\eea
It is important to note that the alternating sum formula (\ref{altsumL}) for $ L ( k,d,n)$ 
does not, by itself, imply the refined formula (\ref{sodsnrefined}). However, the exact sequences 
underlying (\ref{altsumL}), alongside the fact  discussed in Section \ref{sec:exactsequences}
 that the maps in this exact sequence
are $so(d) \times S_n$ invariant, do imply that the sequences can be restricted 
to specific irreps and hence imply the refined counting formulae.

We will now make (\ref{sodsnrefined}) more explicit to produce some general $so(d)\times S_n$ refined counting 
formulae for the space of lowest weight states. 
To determine how many times irrep $\Lambda_{3,2}$ appears in $\Lambda^I(V_{ nat})$, we take
the trace of the projector to  $\Lambda_{3,2}$ from $\Lambda^I(V_{ nat})$. 
The result is
\bea 
&&  \Mult (\Lambda^I (V_{nat}^{(S_n)});\Lambda_{3,2}^{(S_n)})\cr 
&&=\sum_{p\vdash n}\sum_{q\vdash I} {(-1)^{q_2+q_4+\cdots }} {\chi_{\Lambda_{3,2}}^{p}\over Sym(p) Sym(q) }\prod_{i=1}^I \left(\sum_{d|i} d p_d \right )^{ q_i }
\eea
where $\chi_{\Lambda_{3,2}}$ is the $S_n$ character of the permutation with cycle structure $p$ in irrep $\Lambda_{3,2}$
and
\bea
Sym(p)=\prod_i i^{p_i} p_i!
\eea
with $p_i$ denoting the number of parts in partition $p$ that are equal to $i$. 
This is obtained by setting $ \Lambda_2 = [ 1^k] $ in (\ref{refinedSnSkVH}) of Appendix \ref{SnSkrefinedVHk}, and using 
the fact that characters in the anti-symmetric are given by $ (-1)^{q_2+q_4+\cdots } $. 
We now need to consider refining $\Sym^k (V_{ dH})$. 
For the $so(d)$ part, we can use $ so(d)$ characters. 
For $d=3$, we have $so(3)$, so that we only need well known $su(2)$ results. 
For $ d=4$, we will use $ so(4) = su(2) \times su(2)$, so again we need only $su(2)$ results.

\subsection{ Refined Counting: general  $d$ } 

We can  make the formula (\ref{sodsnrefined}) more explicit for general $d$. 
The formulae we write here will be not be as computationally efficient as for $ d=3,4$
but may still be useful in further studies of $so(d,2)$ representations and free field primaries
in higher dimensions. 
We focus on the $d$-dependent quantity 
$$ 
\Mult (  \Sym^{ k  }  ( V_{ dH } ) ;  \Lambda_1^{ so(d)} \otimes  \Lambda_{3}^{ (S_n)} )
$$ 
in (\ref{sodsnrefined}).  $ \Sym^k ( V_{ dH} ) $ is the $S_k$ invariant 
part of $ ( V_d \otimes V_H )^{ \otimes k } = V_d^{ \otimes k } \otimes V_H^{ \otimes k }  $. 
We have the decompositions
\bea 
V_d^{ \otimes k } = \bigoplus_{ \Lambda_1 , \Lambda_2}
 V^{ so(d)}_{ \Lambda_1 } \otimes V_{ \Lambda_2 }^{ (S_k)} \otimes V_{ \Lambda_1 , \Lambda_2 }  
\eea
and
\bea 
V_{ H}^{ \otimes k } = \bigoplus_{ \Lambda_3 , \Lambda_4 } V_{ \Lambda_3}^{ (S_n)} \otimes V_{ \Lambda_4}^{ (S_k)} \otimes V_{ \Lambda_3, \Lambda_4} 
\eea
To count $S_k$ invariants in $V_{dH}^{ \otimes k}$, we use the above while setting $ \Lambda_4 = \Lambda_2 $. This leads to 
\bea 
&& \Mult ( \Sym^k ( V_{dH } , \Lambda_1^{ so(d)} \otimes \Lambda_3^{ (S_n)}  )
= \sum_{ \Lambda_2 \vdash k } \Mult ( V_d^{ \otimes k } , \Lambda_1^{ so(d) } \otimes \Lambda_2^{ (S_k)} )  
\Mult ( V_H^{ \otimes k } , \Lambda_3^{ (S_n)} \otimes \Lambda_2^{ (S_k)  } ) \cr 
&& 
\eea
The  $  \Mult ( V_d^{ \otimes k } , \Lambda_1^{ so(d) } \otimes \Lambda_2^{ (S_k)} )   $
can be calculated using characters of $so(d) $ and $S_k$. 
\bea 
&& \Mult ( V_d^{ \otimes k } , V_{ \Lambda_1}^{ so(d)} \otimes \Lambda_2^{ (S_k) } ) \cr 
&& = { 1 \over k! } \sum_{ \sigma \in S_k } \int dU \chi_{ \Lambda_1 } ( U ) 
\chi_{ \Lambda_2 } ( \sigma ) tr_{ V_d^{ \otimes k } } ( U^{ \otimes k } \sigma ) \cr 
&& =  \sum_{ p \vdash k } \int dU \chi_{ \Lambda_1 } ( U )  
{ \chi_{ \Lambda_2}^p \over Sym ~  p } \prod_{ i }  ( tr U^i )^{ p_i } 
\eea

For the cases of $d=3,4$ we will  give expressions below in terms of generating functions,
 which are more explicit than the group  integrals above. 

\subsection{ Refined counting: $d=3$ case } 

We need the multiplicities of $V_{ \Lambda_1^{so(3)}}\otimes V_{\Lambda_2^{(s_k)}}$ in $V_{3}^{\otimes k}$.
This problem has been considered in \cite{King}. Our coordinates $X_\mu^A$ are in the $3$ of $so(3)$, which is the spin $1$ of $SU(2)$. For $d=3$, $ \Lambda_1$ is parameterised by one integer $l$ for the spin. 
Our multiplicities are given by formula 6.2 of \cite{King}, with $m=2$ for spin $1$. The result is
\bea 
&& \Mult (V_{3}^{\otimes k},[l]\otimes\Lambda_2) \cr
&&= {\rm Coefficient}\left (q^0,(1-q)q^{\sum_{i}c_i(c_i-1)/2  ~ + ~ l / 2 - k  } \prod_{ ( i , j ) \in \Lambda_2 } 
{ ( 1 - q^{ 3 - i + j  } ) \over ( 1 - q^{ h ( i , j ) } ) }  \right )  \cr
&&
\eea
where the notation is an instruction to pick up the coefficient of $q^0$ in the expansion of the second argument above.
$c_i$ is the length of the $i$'th  column of $ \Lambda_2$. $k$ is the number of boxes in $\Lambda_2$. 
$(i,j) $ label the row and column of the boxes in $ \Lambda_2$ and $h(i,j)$ is the hook length of the box. 
The multiplicity $ \Mult (  \Sym^{k - 2I }  ( V_{ dH } ) ;  ( \Lambda_1^{ so(d)} = [ l ] )  \otimes  \Lambda_{3,1}^{ (S_n)} ) ) $ which appears  in the present $so(3)$ instance of (\ref{altsumL}) can be made more explicit. We use  $V_{dH} = V_{ d}^{ so(d)}  \otimes V_H^{ S_n} $ so that the $k$ 
fold tensor power is
\bea 
 (V_{dH} )^{ \otimes k -2I } =  ( V_d^{ \otimes k -2I } \otimes V_H^{ \otimes k -2I } )
\eea
We can decompose the $so(d)$ and $S_n$ parts separately into irreps 
of $so(d) \times S_k $ and $ S_n \times S_k$ respectively. Identifying the $S_k$ irreps and 
summing projects to the invariant  of $S_k$. The outcome is 
\bea 
 && \Mult (  \Sym^{k - 2I }  ( V_{dH}) ;  ( \Lambda_1^{ so(d)} = [ l ] )  \otimes  \Lambda_{3,1}^{ (S_n)} ) )  \cr 
&&  = \sum_{ \Lambda_2 \vdash k - 2I } 
  \Mult ( V_{ 3}^{ \otimes k - 2 I  } ,  [ l ] \otimes \Lambda_2 )  )  
   \Mult (  V_{ H }^{ \otimes k - 2I }  , V_{ \Lambda_{ 3,1} }^{ ( S_n ) } \otimes
    V_{\Lambda_2  }^{ ( S_{ k - 2I } ) }   )  \cr 
   && 
\eea
where 
\bea\label{refinedSnSkVH}
&& \Mult (  V_{ H }^{ \otimes k - 2I }  , V_{ \Lambda_{ 3,1} }^{ ( S_n ) } \otimes
    V_{\Lambda_2  }^{ ( S_{ k - 2I } ) }   )
= \sum_{ p \vdash n } \sum_{ q \vdash k }{ \chi^p_{ \Lambda_1 }  \chi^q_{ \Lambda_2} 
  \over Sym(p) Sym(q)}  \prod_{i=1 }^k \left (    -1 + \sum_{ d | i  } d  p_d              \right )^{ q_i }\cr
&&
\eea
We have again used Appendix \ref{SnSkrefinedVHk}. 
Explicit counting results obtained by implementing the formulas of this section in Sage are given in Appendix \ref{d3count}.

\subsection{ Refined counting : the $d=4$ case } 

The fundamental of $SO(4)$ is the $V_{1/2}\otimes \bar V_{1/2}$ of $SU_L(2) \times SU_R(2)$, where $V_{1/2}$ is the two-dimensional spin half irrep of $SU(2)$. $ X_{ \mu }^{ A} $ transforming 
in $V_4  \otimes V_H$ can be written as $X_{ \alpha , \dot \alpha }^A$ to reflect the description as 
$ V_{1/2} \otimes V_{ 1/2} \otimes V_H$. It is useful to think of $V_{1/2}$ as a two-dimensional rep of $U(2)$, which we call $U_L(2)$ and likewise $ \bar V_{ 1/2} $ as an irrep of $U_R(2)$.  

The decomposition relevant for our discussion is
\bea 
V_{1/2}^{\otimes k} = \bigoplus_{\Lambda_{2,1}\in Y_{2}^{(k)}}V_{\Lambda_{2,1}}^{u_L(2)}  \otimes V_{\Lambda_{2,1}}^{(S_k)} 
\eea
where $Y^{(k)}_{2} $ is the set of Young diagrams with $k$ boxes and at most two rows. 
This decomposition is an example of Schur-Weyl duality (see for example \cite{FultonHarris}). 
The $ u(2)$ irrep associated with a Young  diagram having row lengths $ ( r_1 , r_2) $ has $su(2)$ spin 
$ (r_1-r_2) /2 $ and dimension $r_1-r_2+1$. Similarly we have 
\bea 
\bar V_{ 1/2}^{ \otimes k } = \bigoplus_{ \Lambda_{ 2,2} \in Y_{ 2}^{ (k) } } V_{ \Lambda_{2,2} }^{ u_R (2)}  \otimes V_{ \Lambda_{2,2} }^{ (S_k) } 
\eea
For the $k$'th power of the hook we have 
\bea 
V_H^{ \otimes k } = \bigoplus_{ \substack{ \Lambda_3 \in Y^{ (n)} \\ \Lambda_{2,3}  \in Y^{(k)}    }  } V^{ (S_n)}_{ \Lambda_3 } \otimes V_{ \Lambda_{ 2,3} }^{ ( S_k)}   \otimes V_{ \Lambda_{ 3 } , \Lambda_{2,3} }    
\eea
The dimension of $V_{ \Lambda_{ 3 } , \Lambda_{2,3} }  $  is the multiplicity of the irrep 
$  V^{ (S_n)}_{ \Lambda_3 } \otimes V_{ \Lambda_{ 2,3}}^{ ( S_k)}$ in the tensor product. 
The dimension is given by equation (\ref{refinedSnSkVH}). The final result is
\bea 
&& \Sym^k ( V_{ 4H} ) 
= \bigoplus_{\substack{ \Lambda_3 \in Y^{(n)}  \\ \Lambda_{ 2,1} \in Y_{ 2}^{ (k) } \\ 
 \Lambda_{ 2,2} \in Y_{ 2}^{ (k) }\\\Lambda_{ 2,3} \in Y_{ 2}^{ (k) }   }} V_{ \Lambda_{2,1} }^{ u_L(2)} \otimes V_{ \Lambda_{2,2} }^{ u_R (2)}  \otimes V_{ \Lambda_3}^{ ( S_n)} \otimes V_{ \Lambda_3 , \Lambda_{ 2,3} } \otimes V_{ \Lambda_{ 2,1} , \Lambda_{ 2,2}  ,\Lambda_{ 2,3} }^{ S_k {\rm invts} }  \cr 
 && 
\eea
where $V_{ \Lambda_{ 2,1} , \Lambda_{ 2,2}  ,\Lambda_{ 2,3} }^{ S_k {\rm invts} }$ is the space of $S_k$ invariants
in the Kronecker product $\Lambda_{2,1}\otimes\Lambda_{2,2}\otimes\Lambda_{2,3}$ of $S_k$ irreps.
The multiplicity of irrep $  \Lambda_3^{(S_n)},(l_1,l_2)_{so(4)}$ appearing in $\Sym^k (V_{4H})$ is thus
\bea 
&& \sum_{ \Lambda_{ 2,1} \in  Y^{ ( k)}_2  } \sum_{ \Lambda_{ 2,2 } \in  Y^{ ( k)}_2  }
\sum_{ \Lambda_{ 2,3 } \in Y^{ ( k)} } 
\delta ( r_1 ( \Lambda_{ 2,1} )  - r_2 ( \Lambda_{ 2,1} ) , l_1  )
\delta (   r_1 ( \Lambda_{ 2,2} )  - r_2 ( \Lambda_{ 2,2} ) , l_2  )   \cr
&& \hspace*{.5cm}  \Mult ( V_{ \Lambda_{ 2,3} }^{ (S_k)  }  \otimes V_{\Lambda_3  }^{ (S_n) }  , V_H^{ \otimes k  } )  C ( \Lambda_{ 2,1} , \Lambda_{ 2,2} , \Lambda_{ 2,3} ) 
\eea 
$C$ is the Kronecker coefficient for $ S_k$. 
We can plug this into (\ref{sodsnrefined}) in order to get the refined counting for $d=4$.  
Counting results obtained from these formulas using Sage are given in Appendix \ref{d4count}.

\section{ The Confluent Binomial Transform  and construction }
\label{sec:CBT} 

The exact sequences we derived in Section \ref{sec:exactsequences} have led to a  dimension formula for $\cL (k,d,n)$  (or for $\cI (k,d,n)$)  as an alternating sum involving exterior powers of $V_Q$. In this section we will show that there is a dimension formula for  $\cI (k,d,n)$ as a positive sum involving symmetric powers of $ V_Q$. The two formulas are related by an 
 identity involving binomial coefficients.  There is some superficial similarity to 
equations involved in the binomial transform of combinatorics, but the identity at hand is different. 
As we will explain, the key identity which makes it work is a property of the Tricomi Confluent Hypergeometric Function. Consequently, we name it the Confluent Binomial transform (CBT). 
In this section we will develop these ideas discussing the dimension formula for  $\cI (k,d,n)$ as a positive sum in detail. This  forms the foundation for a construction algorithm for the LWPs.
For this reason, we will refer to the positive  dimension formula in terms of a positive sum as the construction formula. To go beyond counting and get the construction algorithm for the LWPs  requires 
a discussion of an inner product.

\subsection{ From resolution to construction : counting without signs } 

Using characters we have obtained the generating function of the number of LWPs as follows
\bea 
{(1-s^2)^n\over (1-s)^{d(n-1)}} = \sum_{l =0}^{ \infty } L(l,d,n) s^{ l }
\eea
We will derive an interesting expression for $L(l,d,n)$ in terms of $S(k,d,n)$ which is the dimension of $\Sym^k (V_{dH})$.
Our starting point is the explicit expression
\bea 
&& S ( k , d , n ) = { ( ( n-1) d + k -1 ) ! \over k! ( ( n-1) d -1 ) !  } \equiv S ( k , D = d ( n-1) ) \cr
&&  
\eea
The second equality emphasizes the fact that this depends only on $ D = d ( n-1)$. 
Observe that 
\bea 
{ 1 \over ( 1 - s)^D } = \sum_{ k =0}^{ \infty } S ( k , D ) s^k 
\eea
or 
\bea 
{ 1 \over ( 1 - s)^{ d ( n-1) }   } = \sum_{ k =0}^{ \infty } S ( k , d, n ) s^k 
\eea
Then we have
\bea 
{ ( 1 - s^2)^n \over ( 1 - s )^{ d ( n-1) } }
&& = \sum_{ p =0}^{ \infty } \sum_{ k=0}^{ n } S ( p , d , n )  { n! (-1)^k \over k! ( n-k) ! } s^{ p+ 2k }  \cr 
&& = \sum_{ l =0}^{ \infty } \sum_{ k=0}^{ n } S ( l-2k  , d , n )  { n! (-1)^k \over k! ( n-k) ! } s^{ l }
\eea
We now find
\bea 
L ( l , d , n ) = \sum_{ k =0}^{ n } { (-1)^k n! \over k! ( n-k)! } S ( l - 2k  , d, n ) \label{L(S)}
\eea
which is precisely the dimension formula that is implied by the exact sequences described in Section \ref{sec:exactsequences}.
We will now argue that there is a second formula relating $L(l,d,n)$ and $S(l-2k,d,n)$, given by
\bea\label{Sformula} 
S(p,d,n)=\sum_{i=0}^{\lfloor{p\over 2}\rfloor}L(p-2i,d,n){(n+i-1)!\over i!(n-1)!} \label{S(L)}
\eea
Start by substituting the formula (\ref{L(S)}) for $ L( k , d , n ) $ in terms of $S ( k , d, n ) $ into (\ref{S(L)}) to find 
\bea 
S ( p , d, n )  && =  \sum_{ i = 0}^{ \lfloor { p \over 2 } \rfloor   }  \sum_{ k =0 }^{  n } {  (-1)^k n! \over k! ( n-k)! }   { ( n+ i - 1 )! \over i! ( n-1) ! } S ( p - 2i - 2k , d, n ) \cr 
&& = \sum_{ m =0}^{  \lfloor { p \over 2 } \rfloor  } \sum_{ k = 0 }^{ \min  ( m , n ) }  { (-1)^k n \over ( m-k)! k! ( n-k) ! } S ( p - 2m , d, n ) 
\eea
This is indeed an equality, which follows after using the identity  
\bea\label{SBT} 
\sum_{k=0}^{\min (m,n)}{(-1)^k \over (m-k)! k!(n-k)!}={1\over n}\delta_{m,0} 
\eea
The equation (\ref{Sformula}) implies, after subtracting the $i=0$ term $ L ( p , d, n ) $,
that the dimension of the ideal generated by $V_Q$ is 
\bea\label{dimIdeal} 
\Dim ( \cI ( p , d, n ) ) = \sum_{i=1}^{\lfloor{p\over 2}\rfloor}L(p-2i,d,n){(n+i-1)!\over i!(n-1)!}  
\eea

It turns out that the  identity (\ref{SBT}) is related to a hypergeometric function. 
Introduce the function
\bea 
F ( x  ; m , n )  =  \sum_{ m =0}^{  \lfloor { p \over 2 } \rfloor  } \sum_{ k = 0 }^{ \min  ( m , n ) }  { (x)^k  \over ( m-k)! k! ( n-k) ! }  
\eea
With the help of Mathematica, we find 
\bea 
F ( x ; m , n ) =  { (-1)^m x^{ m } \over m! n! } U [ - m , 1 - m + n , -x^{-1} ] 
\eea
where $U$ is a tricomi confluent hypergeometric function. 
Consequently we have
\bea 
\sum_{ m =0}^{  \lfloor { p \over 2 } \rfloor  } \sum_{ k = 0 }^{ \min  ( m , n ) }  { (x)^k  \over ( m-k)! k! ( n-k) ! } =  { (-1)^m x^{ m } \over m! n! } U [ - m , 1 - m + n , -x^{-1} ]
\eea
This reduces our identity to a property of the tricomi confluent hypergeometric function when the last argument is $1$
\bea 
{ 1 \over m! n! } U [ - m , 1 - m + n , 1  ] = { 1 \over n } \delta_{ m , 0 } 
\eea

The equation (\ref{dimIdeal}) gives the dimension of the ideal at each $k$, as a sum of positive terms. 
The ideal generated by the quadratic polynomials $ \{ Q_A : 0 \le A \le n -1  \} $ consists of expressions of the form 
$ \sum_{ A } h_A Q_A$ where $ h_A \in \cR = \mC [ X_{ \mu}^A ] $. 
We can organize the ideal, as a vector space over $ \mC$,   according to how many $Q$'s they contain
if we restrict the coefficients  $h_A$ to be  without $Q$'s, in other words to belong to the 
quotient space $ \cR / \cI$. 
Elements of degree $k$  containing a single $Q$, for example, are of the form 
\bea 
\sum_{ A } l_A Q_A 
\eea
where $ l_A $ is  in $ \cL ( k -2 , d , n ) $, the space of LWPs of degree $k-2$. Consequently, 
a subspace of $ \cI ( k , d , n  ) $ is 
\bea 
 \cL  ( k -2 , d , n ) \otimes  V_Q 
\eea
At this point, we use the identity (\ref{S(L)}), derived with the help of the confluent binomial transform, to decompose 
$\cI (k,d,n)$ as 
\bea 
 \cI ( k , d , n ) = \bigoplus_{ i =1 }^{ \lfloor{ k \over 2} \rfloor  }    \cL ( k - 2 i  , d , n ) \otimes \Sym^i ( V_Q )\label{forI} 
\eea
This decomposition gives a way of constructing the space of LWPs, recursively in $k$. 
Start with $ k = 2$. 
\bea 
\Sym^2 ( V_{dH} ) = V_Q \oplus \cL ( 2, d , n ) 
\eea
From this we read off the fact that the ideal $\cI (2,d,n)$ is $V_Q$ and $\cL (2,d,n)$ is the  complement to $ V_Q$. With an appropriate inner product, to be discussed in the next section, this will be 
an {\it orthogonal}  complement. 
Now use (\ref{forI}) to write
\bea 
\cI ( 4 , d , n ) =  ( \cL ( 2 , d , n ) \otimes V_Q )  \oplus \Sym^2 ( V_Q ) 
\eea
We know $ \cL ( 2, d, n ) $ from the first step so we can construct this.
We then we take the orthogonal complement to $\cI (4,d,n)$ in $ \Sym^4 (V_{dH})$ to obtain $ \cL ( 4 , d , n ) $. 
We then repeat the process: from (\ref{forI}) we can construct $ \cI ( 6 , d , n ) $ using 
\bea 
\cI ( 6 , d , n )  = ( \cL ( 4 , d , n ) \otimes V_Q )  \oplus ( \cL ( 2 , d , n ) \otimes \Sym^2 ( V_Q )
\oplus \Sym^3 ( V_Q ) )\cr
\eea
Now take the orthogonal complement to $ \cI ( 6 , d, n ) $ in $ \Sym^{ 6 } ( V_{dH} ) $ and we get $ \cL ( 6 , d, n ) $. 
A similar construction starting from $ \cL ( 1 , d, n )  = V_{dH} $ will give a recursive construction for all the odd $ k $ cases. 

Much as the exact sequences of vector spaces over $ \mC$ are related to 
exact sequences of modules over $ \cR = \mC [ X_{\mu}^A ]  $, the above equations decomposing 
the polynomials in $ X_{ \mu}^A $ 
of each degree $k$, can be collected into a statement about the ring $\cR$. The quadratic  $Q_A$'s span the vector space $V_Q$. The symmetric algebra of $V_Q$, denoted by $ \Sym ( V_Q ) $ is the  direct sum 
of symmetrised tensor products of all degrees
\bea 
\Sym ( V_Q )  = \bigoplus_{ k=0}^{ \infty } \Sym^k_{\mC} ( V_Q ) 
\eea
The degree $0$ part is defined as $ \mC$. The above decompositions of $ \cR$ at each degree are captured by 
\bea\label{proposal}
\cR  ( d, n ) = \cR ( d , n ) / \cI ( d , n ) \otimes_{ \mC}  \Sym_{ \mC }  ( V_Q ) 
\eea
which indeed follows from the fact that $ \cI$ is generated by quadratic constraints spanning $V_Q$. 
The subscript $ \mC $ indicates that we are tensoring over the base field, not over the ring.

\subsection{ Implementing the construction using the natural inner product  }  \label{natinner}

In the last section we have described an algorithm for the construction of the polynomials that correspond to primary operators. The algorithm works by recursively proceeding in degree, using orthogonality to construct the higher degree spaces from the lower degree ones. The only missing ingredient in the algorithm was the question of which inner product should be used. In this section we will fill this hole and give a detailed account of the algorithms that have been implemented using Mathematica.

Recall that the LWPs are both translation invariant and harmonic.
The requirement of translation invariance is the primary constraint, written using the differential 
operator realization of the conformal group which sets $K_{\mu}={\partial\over\partial x^\mu}$.
The condition that the polynomials are harmonic follows from the equation of motion for the free scalar.
The irrep $V_+$ corresponding to states of the free scalar field, is the space spanned by harmonic and translation
invariant polynomials in $x_{\mu}$. 
To deal with polynomials that correspond to primaries built using a product of $n$ scalar fields, we replace
$x_\mu\to x^I_\mu$ with $I=1,2,\cdots,n$.
The LWPs are now given by the solution to a system of differential equations involving COM conditions 
($d$ equations) and the Laplacian conditions ($n$ equations). 

Our construction makes use of a natural inner product on polynomials in $x_{\mu}$, defined by 
\bea 
< x_{ \mu_1 } \cdots x_{ \mu_k } , x_{\nu_1} \cdots x_{ \nu_k } > 
=  { 1 \over k! } \sum_{ \sigma \in S_k } 
\delta_{ \mu_1 , \nu_{ \sigma (1) } }\delta_{ \mu_2 , \nu_{ \sigma (2) } } 
\cdots \delta_{ \mu_k , \nu_{ \sigma (k ) } }
\eea
Polynomials of different degree are orthogonal.  
There is an obvious extension to polynomials in the multi-particle system, i.e. polynomials of degree $k$ in $x_{\mu}^I$
as follows
\bea 
< x_{\mu_1}^{I_1} \cdots x_{\mu_k}^{I_k} , x_{\nu_1}^{J_1} \cdots x_{\nu_k}^{J_k} > 
=  {1\over k!} \sum_{\sigma\in S_k} \delta_{\mu_1,\nu_{ \sigma (1)}}\delta^{I_1J_{\sigma(1)}}
\cdots \delta_{\mu_k ,\nu_{\sigma (k)}}\delta^{I_k J_{\sigma (k)}}
\eea

The construction starts by recognizing that harmonic polynomials can be obtained as an orthogonal subspace, with
orthogonality given by the above inner product.
To make the argument, start by noting that polynomials of a fixed degree $k$  in $x_{\mu}$ correspond to symmetric 
tensors of degree $k$. The symmetric tensors span the space 
\bea 
\Sym^k ( V_d ) 
\eea
We will now argue that symmetric tensors with non-vanishing trace form a vector subspace of $\Sym^k (V_d)$ which 
is orthogonal to the traceless tensors. 
The traceless tensors correspond to harmonic polynomials, which establishes the result. 

For simplicity, start with the single particle case.
Consider the  differential operator 
$\sum_{\mu,\nu=1}^d x_{\mu} x_{\mu}{\partial^2 \over\partial x_{\nu}\partial x_{\nu}}$. Let it act on all polynomials of fixed degree - it is a linear operator on this space. This linear operator is hermitian with respect to the natural inner product introduced above and hence it is diagonalisable.  
The harmonic polynomials are the null eigenvectors, while the traceful eigenvectors belong to the non-zero eigenspace. 
Since the linear operator is hermitian eigenstates of different eigenvalue are orthogonal with respect to the inner product introduced above. 
This proves that, for the single particle case, the traceless tensors form an orthogonal subspace of $\Sym^k (V_d)$.

The single particle argument is easily generalized: the polynomials in $x_{\mu}^I$ which are annihilated by all $n$ Laplacians must be orthogonal to any polynomial that is not annihilated by one or more Laplacians. 
The hermitian linear operator that plays a role in the multi-particle case is the sum of the single particle
operators
\bea
\cO_\cL\equiv\sum_{I=1}^n \sum_{\mu,\nu=1}^d
x^I_{\mu} x^I_{\mu}{\partial^2 \over\partial x^I_{\nu}\partial x^I_{\nu}}
\eea
Anything harmonic in all the $ x_{\mu}^I$ is in the null space of $\cO_\cL$. 
Any eigenvector not annihilated by all the Laplacians belongs to a non-zero eigenspace of $\cO_\cL$.
Since eigenfunctions with distinct eigenvalues are orthogonal this proves that the multi-harmonic polynomials are orthogonal to the polynomials which are not multi-harmonic. 

We still need to consider the $K_{\mu} $ conditions which we have named the COM conditions above.
A polynomial of degree $k$ in the $ x_{\mu}^I$ is an element of  $\Sym^k (V_{ d} \otimes V_{ nat })$.
We will first outline the argument for the simplest case of $d=1$. 
Start from the observation that $\Sym^k ( V_{ nat } ) = \Sym^k ( V_{0}\oplus V_H)$ which implies the decomposition 
\bea 
\Sym^k ( V_{ nat } ) = \Sym^k ( V_0 \oplus V_H ) 
= \sum_{ l =0}^k \Sym^l ( V_0 ) \otimes \Sym^{ (k-l) } ( V_{ H} ) 
\eea
Since the natural inner product is $S_n$ invariant, this decomposition is orthogonal with respect the natural inner product
\bea 
< x^I , x^J > = \delta^{ I J }  
\eea
The $l= 0$ subspace is annihilated by the COM differential operator. 
Thus, polynomials that obey the center of mass condition again form an orthogonal subspace of $\Sym^k(V_1)$.
A second approach to demonstrate the same fact, makes use of the operator 
\bea
\cO_1 =\sum_{I,J=1}^n x^I{\partial\over\partial x^J}
\eea
which is hermitian with respect to the natural inner product.
The translation invariant polynomials belong to the null space of $\cO_1$, while the non-translation
invariant eigenvectors are not annihilated by $\cO_1$ and hence belong to a non-zero eigenspace of $\cO_1$.
Since eigenfunctions of a hermitian operator with distinct eigenvalues are orthogonal we arrive at our earlier
conclusion that polynomials obeying the center of mass condition again form an orthogonal subspace of $\Sym^k(V_1)$.
As for the Laplacian discussion above, we can generalize this discussion from $d=1$ to general $d$. 
For general $d$ we consider
\bea
\cO_{cm} =\sum_{I,J=1}^n \sum_{\mu=1}^d x^I_\mu{\partial\over\partial x^J_\mu}
\eea
The null space of $\cO_{cm}$ are polynomials invariant under simultaneous translation $x^I_\mu\to x^I_\mu+a_\mu$. 
Anything not annihilated by $\cO_{cm}$ belongs to a non-zero eigenspace of $\cO_{cm}$.
Since eigenfunctions with distinct eigenvalues are orthogonal this proves that the translation invariant polynomials are an orthogonal subspace of $\Sym^k (V_{ d} \otimes V_{ nat })$ for any $d$.

We are now ready to describe our construction algorithm.
The space of polynomials of fixed degree can be decomposed, with respect to 
the hermitian operators $\cO_{cm}$ and $\cO_\cL$ in terms of null and positive eigenvalues as follows 
\bea 
&& \cR (k) =  ( \cR ( k ) )^{ cm}_{0 }  \oplus ( \cR ( k ) )^{ cm}_{ + }   \cr 
&& \cR ( k ) =  ( \cR ( k ) )^{\cL}_{0 }  \oplus ( \cR ( k ) )^{\cL}_{ + }
\eea
This decomposition is orthogonal with respect to the natural inner product. 
The LWPs are in 
\bea 
 ( \cR ( k ) )^{ cm}_{0 }  \cap ( \cR ( k ) )^{\cL}_{0 }  
\eea
In the form just described, it is straight forward to produce a Mathematica implementation of the construction
algorithm that gives the full set of LWPs.
Run times increase as the degree $k$ is increased.

An alternative construction algorithm that exploits more of the $S_n$ structure of the problem, starts by considering
polynomials in the hook variables $X_{\mu}^A$, $A=1,\cdots,n-1$. The advantage is that these polynomials
already satisfy the COM condition, so we only need to impose (\ref{Qdef0}) and (\ref{kappacons}).
Again, by using Mathematica to pick up the subspace orthogonal to (\ref{Qdef0}) and (\ref{kappacons}) we have 
also implemented this alternative construction algorithm.
The above discussion implies that the space of degree $k$ polynomials can be decomposed as 
\bea\label{orthogdec}
p^{(k)} = p_h^{(k)} + Q_A p_{h,A}^{ (k-2) } + Q_AQ_B p_{h,AB}^{(k-4)} + \cdots 
\eea
$p_h^{(k)},p_{h,A}^{(k)},\cdots$ are all polynomials of degree $k$ which are annihilated by the 
Laplacians $ \Box_A$ defined in (\ref{LapH}). 
In the expansion only the first term $p_{ h}^{(k)}$ is in $(\cR (k))^{\cL}_{0}$. 
The remaining terms belong to  $(\cR (k))^{\cL}_{+} $. 
In the expansion of $p^{(k)}$, the leading term $p_h^{(k)}$ is orthogonal 
to all of the subsequent terms.

\subsection{Commutative  star product on lowest weight polynomials }\label{commstar}

As we saw in Section \ref{RingDescription} the lowest weight polynomials are in 1-1 correspondence with a quotient ring, which 
has an associative product inherited from the quotient construction. 
Since the Laplacian constraints obeyed by the polynomials of the ring are second order differential operators, given two polynomials that obey the Laplacian constraints, the product of the two will, in general, fail to. 
Using  (\ref{orthogdec}) we will show there exists a suitable commutative star product so that given two polynomials that obey the 
Laplacian constraints, the star product of the two also obeys the constraints.

The components in the decomposition (\ref{orthogdec}) can be organized by the grading defined by counting the
number of $Q$s.  $Q_A p_{h,A}^{ (k-2)}$ is degree 1, $Q_AQ_B p_{h,AB}^{(k-4)}$ degree 2 and so on.
Polynomials obeying the Laplacian constraints are degree zero.
The key idea behind the star product is that the degree just defined is additive: if polynomial $f_1$ is of degree $k_1$ 
and $f_2$ is of degree $k_2$ then the product $f_1 f_2$ is of degree $k$ with $k\ge k_1+k_2$.
This is almost obvious: consider the product of a degree $k_1$ and degree $k_2$ term
\bea
Q_{A_1}\cdots Q_{A_{k_1}} p_{h,A_1\cdots A_{k_1}}^{(q_1)}
Q_{B_1}\cdots Q_{B_{k_2}} p_{h,B_1\cdots B_{k_2}}^{(q_2)}\label{prodoftwo}
\eea
As we explained above, the product $p_{h,A_1\cdots A_{k_1}}^{(q_1)}p_{h,B_1\cdots B_{k_2}}^{(q_2)}$ will not in general obey the Laplacian constraints. Consequently we can again use the decomposition (\ref{orthogdec}) to write
\bea
p_{h,A_1\cdots A_{k_1}}^{(q_1)}p_{h,B_1\cdots B_{k_2}}^{(q_2)}=
p_{h,A_1\cdots A_{k_1}B_1\cdots B_{k_2}}^{(q_1+q_2)}
+Q_C p_{h,A_1\cdots A_{k_1}B_1\cdots B_{k_2}C}^{(q_1+q_2-2)}+\cdots
\eea
Inserting this back into (\ref{prodoftwo}) proves the result.

With this observation, we can define the star product.
Using (\ref{orthogdec}) decompose the product of two polynomials, which each obey the Laplacian constraints
\bea
f^{(k_1)}_h g^{(k_2)}_h&=&(fg)^{(k_1+k_2)}_h+Q_C (fg)^{k_1+k_2-2}_{h,C}+\cdots\cr\cr
&+&Q_{C_1}\cdots Q_{C_m}(fg)^{(k_1+k_2-2m)}_{C_1\cdots C_m}
\eea
The star product we want is
\bea
f^{(k_1)}_h * g^{(k_2)}_h=(fg)^{(k_1+k_2)}_h
\eea
Only the degree zero term survives because all higher degree terms are set to zero by the ideal of the ring. 

We will now argue that this star product is associative.
Recall that the usual product on polynomials is associative
\bea
f^{(k_1)}_h(g^{(k_2)}_h h^{(k_3)}_h)=(f^{(k_1)}_hg^{(k_2)}_h)h^{(k_3)}_h
\eea
Refining both sides of this last equation according to degree, we can write this as
\bea
&& f^{(k_1)}_h \left((gh)^{(k_2+k_3)}_h+Q_A (g h)^{(k_2+k_3-2)}_{h,A}+\dots\right) \cr
&&\qquad\qquad=((fg)^{(k_1+k_2)}_h
+Q_A (fg)^{(k_1+k_2-2)}_{h,A}+\dots)h^{(k_3)}_h\cr\cr
&&(f (gh)^{(k_2+k_3)}_h)^{(k_1+k_2+k_3)}_h+ Q_A (f (gh)^{(k_2+k_3)}_h)^{(k_1+k_2+k_3-2)}_{h,A}+\dots  \cr
&&\qquad\qquad= ((fg)^{(k_1+k_2)}_h h)^{(k_1+k_2+k_3)}_h
+Q_A ((fg)^{(k_1+k_2)}_hh)^{(k_1+k_2+k_3-2)}_{h,A} +\dots\cr
&&
\eea
Equating degree zero pieces on the two sides we have
\bea
(f (gh)^{(k_2+k_3)}_h)^{(k_1+k_2+k_3)}_h = ((fg)^{(k_1+k_2)}_h h)^{(k_1+k_2+k_3)}_h
\eea
From our definition of the star product we have
\bea
   f*(g*h)=f*(gh)^{(k_2+k_3)}_h=(f (gh)^{(k_2+k_3)}_h)^{(k_1+k_2+k_3)}_h
\eea
\bea
  (f*g)*h=(fg)^{(k_1+k_2)}_h *h = ((fg)^{(k_1+k_2)}_h h)^{(k_1+k_2+k_3)}_h
\eea
It is now evident that
\bea
   f*(g*h)=(f*g)*h
\eea
demonstrating that the star product is indeed associative.

It is reasonable to expect (based on the study of special instances of $ n , d $) 
that this associative product can be expressed 
in terms of the ordinary product $fg$, corrected by products of the form 
$ \cO ( f ) \tilde \cO ( g ) $ where $ \cO , \tilde \cO$ are appropriate 
differential operators. Finding the explicit form of these operators in generality would 
be an interesting exercise for the future.

\section{ Further construction methods for lowest weight polynomials }\label{sec:2moremethods} 

 The construction algorithm we gave in the previous section 
 has a recursive nature, and produces all the LWPs of degree up to any chosen maximum $k$.
 At each $k$, it uses orthogonality to elements written in terms of the LWPs at lower $k$. 
 A second, more direct, algorithm 
works at fixed $k$, and implements the differential equations defining LWPs.
 A third 
algorithm works with projectors, and is based on analogies between the  construction of 
LWPs and that of constructing symmetric traceless tensors. All of these algorithms have been 
tested in Mathematica. We give a brief discussion of algebraic geometry methods for 
constructing the quotient ring at hand.

\subsection{ Intersection of Kernels of two differential operators }

In this section we will outline a closely related but distinct construction algorithm. This new
algorithm uses the fact that, as we explained earlier, the space of LWPs can be identified as the common null space of
a set of differential operators. Here we will consider  degree preserving version of the differential operators, and by using positive semi-definiteness properties, reduce the problem to 
that of finding the simultaneous null space of two differential operators. This last step is 
 implemented in 
Mathematica. 

Consider the space of polynomials of degree $k$ in the $dn$ variables $x_{\mu}^{I}  $ where $1\le  I \le n $. 
The degree $k$ is a sum of degrees of $ k = k_1 + k_2 + \cdots + k_n $, where $k_I $ 
is the degree in the $I$'th variable. A general polynomial with specified 
degrees $ ( k_1 , k_2 , \cdots , k_n )  $ is 
\bea 
 X^{\vec k}_{\vec \mu} &=&
X_{ \mu_{11} , \mu_{ 12} ,  \cdots , \mu_{ 1 k_1 } ; \mu_{ 21} , \mu_{22} , \cdots , \mu_{ 2k_2 } ; 
\cdots ; \mu_{ n 1 } , \mu_{n 2 } , \cdots ,\mu_{ n k_n  } }\cr \cr
&=& x_{ \mu_{11}}^{1} \cdots x_{ \mu_{ 1 k_1} }^{1} 
 x_{ \mu_{ 2 1 }}^{2} \cdots x_{ \mu_{ 2 k_2 } }^{2}  
 \cdots x_{ \mu_{ n  1 }}^{n} \cdots x_{ \mu_{ n  k_n  } }^{n} 
\eea
All the $ \mu $ indices take values in the range $ 1 \le \mu \le d$. 
$X$ is symmetric in the first $ k_1 $ indices, the next $ k_2$ indices, etc. 
The number of independent $X$'s is 
\bea 
\prod_{ I =1}^{ n } { ( d + k_I -1 ) ! \over k_I ! (d - 1 ) ! }  
\eea
We will be considering the vector space of the $X$'s, for all $ \vec k$ satisfying $ \sum_{ I } k_I = k $. This is equivalently a sum over partitions of $k $ with up to $n$ parts (since some of the $k_I$ could be zero). 
This vector space, denoted $\cW_{k;n,d}$,  has dimension 
\bea 
\Dim (\cW_{k;n,d}) = 
\sum_{k_1,k_2,\cdots ,k_n =0}^{k}\delta (k,k_1+k_2+\cdots +k_n)\prod_{I=1}^{n} {(d+k_I-1) ! \over k_I ! (d-1)!}  
\eea
\bea 
\cW_{k;n,d} = \bigoplus_{ \substack { k_1 , k_2 \cdots , k_n =0 \\ \sum_{ I } k_I = k  } }^{k}  \Sym^{ k_1 } ( V_d ) \otimes \Sym^{ k_2 } ( V_d ) \otimes \cdots \otimes \Sym^{ k_n } (  V_d )  
\eea
On this subspace we have the linear operators 
\bea
\cO^{ (I)}_{ \cL} = ( x^{I})^2 \Box_{(I)} = \sum_{ \alpha , \beta = 1 }^{d} 
 x^{I}_{ \alpha }  x^{I}_{ \alpha } { \partial \over \partial x^{I}_{ \beta } \partial x^{I}_{ \beta } }
\eea
which are degree preserving versions of the Laplacians.

Consider the sum of  these $n$ operators 
\bea 
\cO_{ \cL  } = \sum_{ I =1}^n  \cO^{( I) }_{ \cL  }  
\eea
The null space of this operator obeys all the Laplacian conditions. 
The different Laplacian operators are commuting operators and they are all 
positive semi-definite operators, i.e. they all have eigenvalues which are non-negative. 
Consequently, the vanishing of the sum guarantees the vanishing  of the summands. 
Thus, polynomials in the null space of $\cO_{ \cL  }$ are harmonic.

For each $ \alpha \in \{ 1, 2 , \cdots , d  \} $, we have a centre of mass operator 
\bea 
{ \partial \over \partial x^{ CM}_{ \alpha } }  = \sum_{ I =1}^{ n } { \partial \over \partial x^{I}_{ \alpha } } 
\eea 
Polynomial functions of $\{ x_{ \alpha }^{  I } :1 \le  I \le  n \} $ can be factored 
into sums of centre of mass-dependent functions times functions of the differences. 
The degree-preserving  COM operator   
\bea 
\sum_\alpha x^{ CM}_{ \alpha } { \partial \over \partial x^{ CM}_{ \alpha } } = 
\sum_\alpha \sum_{ I , J =1}^{ n  }  x^{I}_{ \alpha } { \partial \over \partial x^{J}_{ \alpha } }
\eea 
is positive semi-definite. Any operator of fixed degree, annihilated by
$  { \partial \over \partial x^{ CM}_{ \alpha } } $ is also annihilated by  
$ x^{ CM}_{ \alpha } { \partial \over \partial x^{ CM}_{ \alpha } } $. 
Therefore the following sum 
\bea 
\cO_{ \mathcal{C} \mathcal {M}  } && =  \sum_{ \alpha =1 }^{d} x^{ CM}_{ \alpha } { \partial \over \partial x^{ CM}_{ \alpha } }  \cr 
&& =  \sum_{ I , J =1}^{ n  }  \sum_{ \alpha =1 }^{d} x^{I}_{ \alpha } { \partial \over \partial x^{J}_{ \alpha } } 
\eea
has the property that its null space is the simultaneous null space of all the $d$ COM operators. 

The null space of $ \cO_{\cC \cM} $ obeys the lowest weight condition. 
We build the combined operator 
\bea 
\cO  = \begin{pmatrix}
\cO_{ \cL  }\\ 
\cO_{ \cC \cM } 
\end{pmatrix} 
\eea
$\cO$ is an operator in $ End ( \cW_{ k ; n , d  } ) $. 
The null space of this operator is the space of lowest weight polynomials of 
degree $k$. 
We have implemented this algorithm in Mathematica and have checked that the dimension of the
null space of $\cO$ does indeed agree with the number of LWPs. 

\subsection{ Constraints and Projectors on $V_{dH}\otimes V_{dH}$}\label{sec:kbasis}

An interesting algebraic angle on the primaries problem is that in some sense it is a generalization of the problem of finding symmetric traceless tensors of $SO(d)$. 
Nice bases for these tensors can be constructed using Young diagram techniques for $SO$ groups. 
These symmetric traceless tensors are annihilated by contraction operators. 
The contraction tensors form part of a Brauer algebra, so the symmetric traceless tensors  of rank $k$ 
for $so(d)$ are related to irreps of a Brauer algebra. This follows from the fact that the commutant of $SO(d)$ in 
$V_d^{ \otimes k}$, where $V_d$ is the fundamental of $so(d)$, is the Brauer algebra. 
In the problem of symmetric tensors we are trying to find tensors
\bea 
T_{ \mu_1 , \mu_2 , \cdots , \mu_k }      
\eea
such that 
\bea 
     T_{ \mu , \mu , \mu_3 , \cdots , \mu_k } = 0 
\eea
Note that, because $T$ is symmetric, we can move the paired indices to any slot. 
Another way to phrase this is by considering the contraction operator
\bea 
C_{ 12} T_{ \mu_1 ,\cdots , \mu_k } = \delta_{ \mu_1 , \mu_2}  T_{ \mu , \mu , \mu_3 , \cdots , \mu_k } = 0 
\eea
The contraction operator is a projector from $ V_d \otimes V_d $ to the trivial rep of $ so(d)$. 

In the present case, there is a natural generalized symmetric tensor in the game
\bea 
T_{ \mu_1 , \mu_2 , \cdots , \mu_k }^{ A_1, A_2 , \cdots , A_k } \leftrightarrow X_{\mu_1}^{A_1} X_{\mu_2}^{A_2} \cdots X_{\mu_k}^{ A_k} 
\eea
The symmetry is the $S_k$ group of permutations of the pairs $(\mu,A)$. 
To start it is constructive to consider the $ k=2$ case. 
Define 
\bea 
V_{dH} = V_d \otimes V_H 
\eea
We are looking at the subspace of 
\bea
V_{dH}^{ \otimes 2 } 
\eea
which is invariant under the $S_2$ permutation of the two factors, i.e. we are looking at 
\bea 
\Sym^2 ( V_{dH} ) 
\eea
Now $V_d \otimes V_d$ contains a symmetric rank 2 tensor, an anti-symmetric rank two tensor and a trace (invariant) of $SO(d)$ all with multiplicity $1$. The product $V_H\otimes V_H $ decomposes under the diagonal $S_n$ as 
\bea 
V_H \otimes V_H = V_0 \oplus V_H \oplus V_{ [ n-2,2] } \oplus V_{[ n-2,1,1] } 
\eea
The symmetric part contains the first three spaces in the direct sum 
\bea 
\Sym^2 ( V_H \otimes V_H ) =  V_0 \oplus V_H \oplus V_{ [ n-2,2] }
\eea
The constraints tell us that the projection to $(V_0^{(S_n)}\oplus V_H^{(S_n)}) \otimes V_0^{SO(d)}$  
inside $\Sym^2 (V_{dH})$  vanishes. So we are looking at vectors
\bea 
 v \in \Sym^2 ( V_{dH} ) 
\eea
which obey 
\bea 
 ( P^{ (SO(d))}_{ 0 } ( P_{0    }^{ (S_n )}  + P_{ H}^{ (S_n)} ) ) ~~   v = 0 
\eea
Let us define 
\bea 
P^{ \cL } =  ( P^{ (SO(d)) }_{ 0 } ( P_{0    }^{ (S_n )}  + P_{ H}^{ (S_n)} ) )
\eea
The operator  $ P^{ \cL } $  is a projector obeying $ (P^{ \cL }) ^2 = P^{ \cL }  $ 
which follows from 
\bea
( P^{ (SO(d)}_{ 0 })^2 &=& P^{ (SO(d)) }_{ 0 } \cr 
( P_{0    }^{ (S_n )} )^2 &=& P_{0    }^{ (S_n )} \cr 
( P_{ H}^{ ( S_n) } )^2 &=& P_{ H}^{ ( S_n) } \cr  
P_{0    }^{ (S_n )} P_{ H}^{ ( S_n) } &=& 0 \cr 
P^{ (SO(d)) }_{ 0 }P_{0    }^{ (S_n )} &=& P_{0    }^{ (S_n )} P^{ (SO(d)) }_{ 0 } \cr 
P^{ (SO(d)) }_{ 0 }P_{H    }^{ (S_n )} &=& P_{H    }^{ (S_n ) } P^{ (SO(d)) }_{ 0 } 
\eea

This is a projector $P^{ \cL} $ which acts on pairwise slots. 
It is the analog of the contraction tensor of the Brauer algebra.
Combine the $ \mu, A$ indices into a composite index $M$. 
We are considering symmetric tensors 
\bea
T_{ M_1, \cdots , M_k } 
\eea
that are annihilated by $P^{\cL}$
\bea\label{vanishingproj}  
P^{ \cL }_{ 12}  T_{ M_1, M_2, \cdots  , M_k } 
= P_{ M_1 M_2}^{ N_1 , N_2}  T_{ N_1, N_2 ,M_3 ,  \cdots  , M_k }  =0 
\eea
We have a very concrete formula for $ \kappa_{ A  BC }$ and hence for $P$ 
\bea 
(P)_{ \mu_1, A_1, \mu_2, A_2}^{ \nu_1 , B_1 ; \nu_2 , B_2} 
= \sum_{ A=0}^{ n-1} \delta_{ \mu_1 , \mu_2} \delta_{ \nu_1 , \nu_2} 
\kappa_{ A , A_1 , A_2 } \kappa_{ A , B_1 , B_2 } 
\eea
The symmetry of $T$ then also implies 
\bea 
 P_{ M_1 M_2}^{ N_1 , N_2}  T_{ N_1, M_3 ,N_2 ,  \cdots  , M_k }  =0 
\eea
and so on for any pair. We will now argue that to find a linear basis at degree $k$ in the ring of LWPs, we have to consider 
symmetric tensors $T$  obeying equation (\ref{vanishingproj}).

Use the inner product for polynomials in  $ x_{ \mu}^I $ used before. 
This induces an inner product of the same form on $X_{ \mu}^A$. 
For  operator  
\bea 
\Box_A = \sum_{ \mu =1 }^d  \sum_{ B , C =1}^{ n-1}  \kappa_{ABC} { \partial^2 \over \partial X_{ \mu}^B \partial X_{ \mu}^C } 
\eea
consider 
\bea
 ( \Box_A )^{ \dagger} \Box_A 
= \sum_{B,C,D,E=0}^{ n-1 } 
\sum_{ \mu , \nu =1}^d  X_{ \nu}^D X_{ \nu}^E \kappa_{ADE}\kappa_{ABC} { \partial^2 \over \partial X_{ \mu}^B \partial X_{ \mu}^C }
\eea 
This is a positive semi-definite operator. Any eigenvector $v$ 
 of eigenvalue $ \lambda $ has the property 
\bea 
( v , \Box_A^{ \dagger} \Box_A v ) = \lambda ( v,v) = ( \Box_A v , \Box_A v  )  \ge 0 
\eea
so $ \lambda \ge 0$. Hence being in the simultaneous null spaces is equivalent to being 
in the  null space of 
\bea 
\sum_{ A =0 }^{ n-1}   ( \Box_A )^{ \dagger} \Box_A 
\eea
Symmetric tensors in this null space are equivalently in the null space of $P^{\cL}_{12}$. 

\subsection{ Standard algebraic geometry methods for $ \cR / \cI  $  } 

As explained in Section \ref{RingDescription}, the LWPs are 
in 1-1 correspondence with the elements of $ \cR / \cI $. 
The quotient ring is defined in terms of equivalence classes. Each equivalence class contains 
an LWP. There are standard algebraic geometry methods, based on Groebner bases, 
for the construction of the equivalence classes. The Groebner bases rely on 
choosing certain orderings on monomials  \cite{Cox}. To pick the LWPs within each equivalence class 
would probably be a non-trivial  additional step. The polynomials $Q_A$ are polynomials 
with integer coefficients, so another approach to the quotient ring may be to use 
an analog of Groebner bases which works for rings defined over integers (see e.g. \cite{cayoub}). 
If we are interested in the effective action problem \cite{HLMM17}, it is only the 
quotient ring which is of interest. If we are interested in constructing primary fields in CFT, 
the specific LWPs are important. It will be interesting to investigate 
the efficacy of the different algorithms given here, relative to the algebraic geometry methods, 
from the point of view of primary fields as well as from the point of view of effective actions.

\section{ Discussion and Future Directions.  }\label{sec:future}

We have considered the problem of constructing primary fields in free scalar CFTs in general dimensions,
 combining  insights from \cite{cft4tft2,DRRR17-PRL,DRRR17} and \cite{HLMM17}. 
This has been a fruitful avenue, with the key results 
  described in the introduction and developed in the bulk of the paper.

A number of future projects are suggested by our results. 
We have given a number of Mathematica constructions of lowest weight polynomials.
These codes are available upon request.
 
It would be interesting to compare the efficiency of the different algorithms discussed 
in Sections \ref{sec:CBT} and \ref{sec:2moremethods}.  Extending the present work to 
fermions and gauge fields is a worthwhile avenue. For the holomorphic sector 
of primaries in the free fermion theory see \cite{DRV18}.

To get primary fields in the CFT, we have to project to $S_n$ invariants. 
The Hilbert series of the $S_n$ invariants is easy to write in terms of 
those for the LWPs (see \cite{cft4tft2}).  A complete description of the 
ring structure for  $S_n$ invariants in the $d=3,4$ is a worthwhile goal in the short term.

It would be interesting to investigate for more general rings, 
  the connection between resolution and construction we have given in Section \ref{sec:CBT}.  
  In the present case all the constraints are quadratic.
  There should be closely related generalizations when the constraints are each homogeneous 
  but of different degrees.

\subsection{ Further developing the Analogy to tracelessness : a generalization of Brauer algebras  } 

In section \ref{sec:kbasis}  we developed an approach to the construction of 
LWPs, based on projectors  acting on degree $k$ polynomials 
in $ X_{ \mu}^A$. These polynomials form a vector space isomorphic to the space of
 symmetric tensors $ \Sym^k ( V_{ dH} ) $. 
It is useful to consider the tensor product $ V_{ dH}^{ \otimes  k  } $
where we have the projectors $ \cP^{ \cL }_{ ij} $ acting on the slots labeled $ i  , j $, 
 and subsequently projecting to the $S_k$ symmetric part. 
This is analogous to the problem of constructing  symmetric traceless tensors in 
$V_d^{ \otimes k} $. In this case there are finite algebras, the Brauer algebras $ D_k ( d ) $ 
the commutant of $o(d) $ in $V_d^{ \otimes k } $, which give a representation theory meaning 
to this construction. For the representation theory of $ D_k ( d ) $, see for example 
\cite{RamChars}. In the present case, we have an analog  of 
the Brauer algebra, namely the algebra generated by  $ \cP^{ \cL }_{ ij} \subset End ( V_{ dH}^{ \otimes k  } ) $ along with permutations in $S_k$. Let us call this algebra $ \cA ( k , d , n )$. 
It is plausible that the LWPs form irreducible reps of this algebra. 
It would be interesting to investigate this conjecture. 

As explained in \cite{cft4tft2}, part of the motivation for studying free field primaries 
comes from the goal of finding a uniform framework of algebraic structures, based on 
two dimensional topological field theory (TFT2), for understanding both the space-time dependence
and the combinatoric structure of correlators in $N=4$ SYM. The combinatorics of the 
half-BPS sector is controlled by a Frobenius algebra (TFT2) which is the space 
of conjugacy classes of symmetric groups $S_n$ for all $n$, resulting in useful Young diagram bases 
for AdS/CFT \cite{cjr}. 
For the quarter BPS sector 
at zero coupling, we have an algebra which is a subspace of $ \mC ( S_{ n +m} ) $ invariant
 under conjugation by $ S_{ n } \times S_m $ \cite{QuivCalc,YusukeTFT2,PCA}. 
 This algebra arises as a way to describe 
 nice bases for the ring of polynomial gauge invariant functions of two matrices $Y , Z $, which are orthogonal under free field inner products \cite{BCD08,BDS08}. Analogous constructions  
 based on Brauer algebras provide an alternative  approach to these orthogonal bases \cite{KR1}. 
 Generalizing these 
 constructions to matrix systems forming representations of general symmetry groups 
 led to an initial foray into the problem of constructing primary fields \cite{BHR1,BHR2}. 
 
 The theme of finite dimensional algebras controlling questions about infinite dimensional 
 representations of the conformal group has been a recurring theme. 
 Developing the algebraic approach to LWPs based on representations of 
 $ \cA ( k , d , n )$ would be a concrete manifestation of the unity
 between algebraic structure for combinatorics and space-time dependence of 
 correlators of gauge invariant observables.

\subsection{ Quadratic algebras and  Koszul algebras.  } 

We have shown that the LWPs are in 1-1 correspondence
with the quotient ring $ \cR / \cI$, where $ \cR = \mC  [ X_{ \mu}^A ] $
and $ \cI$ is generated by (\ref{ExpConsI}). 
This is an example of a quadratic algebra. These are defined by quotients of 
the tensor algebra  $ T ( \mathbb{ V } ) $ 
 of a vector space $ \mathbb{ V } $, determined by a quadratic form $ R \subset \mV \otimes \mV $
 \cite{PP}. In the present case, $\mV = V_{ dH} $ (let us use basis vectors $e_{\mu}^A$)  and 
 $R$ is spanned by 
 \bea 
 && e^{A}_{ \mu} \otimes e^{ B}_{ \nu} - e^{ B }_{ \nu} \otimes e^A_{ \mu} , \cr
&& \sum_{ \mu } \sum_{ B , C  }  \kappa_{ ABC} e_{\mu}^{ B } \otimes e_{ \mu}^C 
 \eea 
 The quotient is $ T ( \mV ) / < R > $. For the explicit 
 definition of $<R>$ see Chapter 4 of \cite{WuGang}. It involves tensoring 
 with arbitrary tensor powers of $ \mV$ on the left and right. 
 The first line above ensures that we project from the tensor algebra to the symmetric algebra. 
 Equivalently we go from  generators  of free algebras to commuting generators $ X_{ \mu}^A$. 
 Every quadratic algebra has a dual quadratic algebra defined by $R^{ \perp} \subset V^* \otimes V^* $. 
 In the present case, we can work out that 
 \bea 
&&  R = \Lambda^{ 2 } (  V_{ dH } )  ~~~ \oplus ~~~ \hbox{Invt}_{ so(d) } ( \Sym^2 ( V_d)  ) 
\otimes ( \Sym^2(V_H) )_{[n] + [ n-1,1] } \cr 
&& \cr 
&& R^{ \perp}  = \Lambda^2 ( V_d^*  ) \otimes \Lambda^2 ( V_H^*  ) ~~~
 \oplus ~~~  (\Sym^2 ( V_d^* ) )'  \otimes \Sym^{ 2 } ( V_H^* )  \cr 
&& \hspace*{2cm}   \oplus  ~~~~ ( \hbox{Invt} ( \Sym^2(V_d^* ) ) \otimes ( \Sym^2(V_H^* ) )_{ [n-2,2] } \cr 
&& 
 \eea 
The details are not important. An important observation is that  $ T ( \mV  ) / < R > $ 
is a commutative algebra due to the presence of  $ \Lambda^{ 2 } (  V_{ dH } )  $ 
  as a direct summand in $ R$, while  $ T ( \mV  ) / < R^{ \perp }  > $ is not commutative 
  due to the lack of such a direct summand. 
  
  A special class of quadratic algebras are said to be Koszul, which happens when the 
  algebras form part of certain exact resolutions of the base field (see \cite{PP,WuGang}). 
   Koszul algebras have a property of 
  Koszul duality whereby the quadratic algebra and its dual quadratic algebra have equivalent 
   derived categories of modules ( see \cite{Wiki-Koszul} and references therein).
    An interesting question is whether $ \cR / \cI $ is Koszul. 
  If it is, this will be much more than a mathematical curiosity. 
  It is plausible that modules of $ \cR / \cI $ control the properties of 
  primary fields for fermions, gauge fields and higher spin fields. This is expected  by analogy 
  to non-commutative geometry where the configuration space of a scalar field becomes 
  a non-commutative algebra and the configuration spaces 
   of non-trivial fields becomes modules over the algebra 
  \cite{Szabo}. 
  Thus if the Koszul property holds in the present case, the physics of $ \cR / \cI $ 
  may be equivalent to the physics of the Koszul dual algebra. This would indicate there might exist  a hidden    non-commutative reformulation of  ordinary  quantum field theory !

So is $ \cR / \cI $ a  Koszul algebra? 
Deformations of this algebra of lowest weight primaries, where the coefficients 
$ \kappa_{ ABC} $ are modified so that they  satisfy a genericity condition,  are Koszul
\cite{FL2002}. A useful fact ( Example 2 following Corollary 6.3 in \cite{PP}) 
is that $ \mC [ x_1 , \cdots , x_n ] / <  q_1 , \cdots , q_r > $ is 
Koszul if $ q_1 , q_2 , \cdots , q_r $ form a regular sequence of quadrics. 
Applying this to our case, the question is whether the $\{ Q_0, Q_1 , \cdots , Q_{ n-1} \} $ 
(without deformation to reach the genericity condition of \cite{FL2002}) 
form a regular sequence. We leave this as a question for the  future.

\subsection{Future direction : Coherence relations between two products.   }

 We showed in \cite{cft4tft2} that the OPE in free scalar theory can be used to
 define a commutative $so(4,2)$ covariant algebra with a non-degenerate bilinear pairing.  The crossing equation of CFT 
 becomes ordinary associativity of the algebra. Here we have seen 
 that there is an algebra controlling primary fields for every $n$. 
 The interplay between the commutative algebra coming from the  OPE and the algebra studied 
  here is an interesting question for the future.

\bigskip 

\begin{center} 
{ \bf Acknowledgements}
\end{center} 
This work of RdMK  is supported by the South African Research Chairs
Initiative of the Department of Science and Technology and National Research Foundation
as well as funds received from the National Institute for Theoretical Physics (NITheP).
SR is supported by the STFC consolidated grant ST/L000415/1 “String Theory, Gauge Theory \& Duality” and  a Visiting Professorship at the University of the Witwatersrand, funded by a Simons Foundation grant held at the Mandelstam Institute for Theoretical Physics.  SR thanks the Galileo Galilei Institute for
Theoretical Physics for hospitality and the INFN for partial support
during the completion  of this work.

\vskip3cm

\begin{appendix} 

\section{ The invariant in $ V_H \otimes V_H \otimes V_H $   }\label{sec:invtVH}

In this section we will give the derivation of (\ref{kappaformula}).
Inserting the explicit expressions for the $S_{AI}$ we have
\bea 
&& \sum_{ I } S_{ C I  }  S_{ B I } S_{ A I }    = \cN_{ A } \cN_{ B } \cN_{ C } 
\sum_{ I } \left ( - C ~ \delta_{ I , C +1  }  + \sum_{ J_1  =1}^{ C } \delta_{ J_1  , I}  \right )
\cr 
&& \hspace*{3cm} 
\left ( - B  ~ \delta_{ I , B +1  } + \sum_{ J_2  =1}^{ B } \delta_{ J_2  , I}  \right )
\left ( - A ~ \delta_{ I , A +1  } + \sum_{ J_3  =1}^{ A } \delta_{ J_3  , I}  \right ) \cr 
&& 
\eea
Expanding the brackets out there are $8$ terms.
Call them $T_1, T_2, \cdots T_8$. 
We will deal with each term separately in what follows.
\bea 
T_1  && = - ABC \sum_{ I } \delta_{ I , C+1} \delta_{ I , B+1} \delta_{ I  , A +1 }  \cr 
     && = - ABC \delta_{ A B } \delta_{ B C } \cr 
      && = - ABC  \delta_{ A , B , C } 
\eea
\bea
T_2 && = BC \sum_{ I =1}^n \delta_{ I , C+1 } \delta_{ I , B+1} \sum_{ J_3 = 1 }^A \delta_{ J_3 , I } \cr 
&& = BC \delta_{ B , C } \sum_{ I =1}^n \delta_{ I , B+1} \sum_{ J=1 }^A \delta_{ J , I } \cr 
&& = BC \delta_{ B , C } \sum_{ I = 1 }^A \delta_{ I , B+1}  \cr 
&& = BC \delta_{ B , C } \Theta ( B < A )  
\eea
\bea 
T_3 && =  AC \sum_{ I } \delta_{ I , C +1 } \delta_{ I , A+1} \sum_{ J_2 =1}^B \delta_{J_2,I} \cr 
&& = A C \delta_{ A , C } \sum_{ I =1}^n \delta_{ I , C+1} \sum_{ J_2 =1}^B \delta_{ J_2 , I } \cr 
&& = AC \delta_{ A , C } \sum_{ I =1}^{ B-1} \delta_{ I , B- C } 
\eea
The last delta function is only non-zero if $ B \ge C+ 1$. Define 
\bea 
\Theta ( B > C  )  && = 1   ~~~ \hbox { if } B > C \cr 
                  && =  0 ~~~~ \hbox { otherwise } 
\eea
We can then write 
\bea 
T_3 = AC \delta_{ A ,C } \Theta ( B > C ) 
\eea
For the fourth term, we have  
\bea 
 T_4 && = \sum_{ I =1}^{ n } \delta_{ I , C+1 } \sum_{ J_2 = 1 }^B 
\sum_{ J_3 =1 }^{ A } \delta_{ J_2 , I }  \delta_{ J_3 , I } \cr 
&& = \sum_{ J_2 =1}^{ B } \sum_{ J_3 =1 }^{ A } \delta_{ J_2 , C+1 } \delta_{ J_3 , C +1 } \cr 
&& = \Theta ( B > C ) \Theta ( A > C )   
\eea
Continuing, we have 
\bea 
T_5 && = BA \sum_{ I } \delta_{ I , B+1 } \delta_{ I , A +1} \sum_{ J_1 =1 }^C \delta_{ J_1 , I } \cr 
&& = BA \delta_{ B A } \sum_{ I } \delta_{ I , A +1 } \sum_{ J_1 =1 }^{ C } \delta_{ J_1 , I } \cr 
&& = BA \delta_{ B , A  } \Theta ( A < C ) 
\eea
\bea 
T_6 && = -B \sum_{ I } \delta_{ I , B+1} \sum_{ J_1 =1}^C \sum_{ J_3 =1 }^{ A } \delta_{ J_3 , I } \delta_{ J_1 , I } \cr 
&& =- \sum_{ J_1 =1}^C \sum_{ J_3 =1}^A \delta_{ J_3 , B+1 } \delta_{ J_1 , B+1 } \cr 
&& = -B \Theta ( B < C ) \Theta ( B < A ) 
\eea
\bea 
T_7 && = - A \sum_{ I } \delta_{ I , A +1 } \sum_{ J_1 =1}^C \sum_{ J_2 =1}^B \delta_{ J_1 , I } \delta_{ J_2 , I } \cr 
&& = - A \sum_{ J_1 =1 }^C \sum_{ J_2 =1 }^B \delta_{ J_1 , A+1 } \delta_{ J_2 , A+1 } \cr 
&& = - A \Theta ( A < C ) \Theta ( A < B ) 
\eea
\bea 
T_8 && = \sum_{ I } \sum_{ J_1 =1 }^C \sum_{ J_2 =1}^B \sum_{ J_3 =1 }^{A } 
     \delta_{ J_1 , I } \delta_{ J_2 , I } \delta_{ J_3 , I } \cr 
     && =  \sum_{ J_1 =1 }^{ C }  \sum_{ J_2 =1}^B \sum_{ J_3 =1 }^{ A } \delta_{ J_1 , J_3 } \delta_{ J_2 , J_3 } \cr 
     && = \sum_{ J_1 =1 }^{ C }  \sum_{ J_2 =1}^B \sum_{ J_3 =1 }^{ A } \delta_{ J_1 , J_2 , J_3 } \cr 
      && = \hbox { Min }  ( A , B , C ) 
\eea
Summing these terms gives (\ref{kappaformula}).

\subsection{ The $ \kappa $ polynomial  }

The $3$-index invariant $ \kappa_{ ABC } $ can be used to define a symmetric polynomial in 
 $ z_1 , z_2  , \cdots , z_{ n -1} $. 
 \bea 
&&  \kappa ( z_1  , z_2 , \cdots , z_{ n-1} ) 
 = \sum_{ A, B , C } \kappa_{ A , B , C } z_A z_B z_C \cr  
&& = - \sum_{ A } A^3 z_A^3 + \sum_{ A > B } B^2 z_A z_B^2 +   \sum_{ C > A  } A^2 z_A^2 z_C 
+ \sum_{ A < B } A^2 z_A^2  z_B  - \sum_{ C < A ; C < B } C z_A z_B z_C \cr 
&&  - \sum_{ B < C ; B < A } 
B z_A z_B z_C - \sum_{ A < B ; A < C  } A  z_A z_B z_C
+ \sum_{  A , B , C } Min ( A , B , C ) z_A z_B z_C  \cr 
&& =  - \sum_{ A } A^3 z_A^3 + 3 \sum_{ A <  B } A^2 z_A^2 z_B - 3 \sum_{ A < B , A < C } A  z_A z_B z_C 
+\sum_{  A , B , C } Min ( A , B , C ) z_A z_B z_C \cr 
&& 
 \eea
We used a renaming of summation variables to get the last line. 
The third term can be manipulated by separating into the case $ B = C $, the case $ A < B < C $ and the 
case $ A < C < B $ to give 
\bea
&& - 3 \sum_{ A < B, A < C } A z_A z_B z_C = 
- 3 \sum_{ A < B  } A z_A z_B^2 - 3 \sum_{ A < B < C } A z_A z_B z_C - 3 \sum_{ A < C < B } 
 A z_A z_B z_C \cr 
 && = - 3 \sum_{ A < B  } A z_A z_B^2  -  6 \sum_{ A < B < C } A z_A z_B z_C
\eea
The last term  can be separated into the cases $ A = B = C $, the case where two are equal and smaller than the third, and the case where two are equal and larger than the third, and the case where all are different
\bea\label{KappaPoly}  
&& \sum_{ A , B , C } Min ( z_A , z_B , z_C ) z_A z_B z_C 
= \sum_{ A } A z_A^3 + 3 \sum_{ A < B }A z_A^2 z_B  + 3 \sum_{ A < B } A z_A z_B^2 
+ 6 \sum_{ A < B < C } A z_A z_B z_C \cr 
&& 
\eea
Collecting all the terms leads to some cancellations and simplifications : 
\bea\label{resultKappa} 
\kappa ( z_A ) = \sum_{ A } A ( 1- A^2) z_A^3 + \sum_{ A < B } 3 A ( 1 + A ) z_A^2 z_B  
\eea
In particular there are no terms where all the indices are different.

Consider now the polynomial $ \kappa_A ( z_1 , z_2 , \cdots , z_{ n-1} ) $ 
\bea 
\kappa_{ A } ( z ) = \sum_{ B , C } \kappa_{ A B C } z_B z_C 
\eea
Consider 
\bea 
{ \partial  \kappa ( z )  \over \partial z_a } && =  { \partial \over \partial z_a } \sum_{ A , B , C }  \kappa_{ A B C } z_A z_B z_C  \cr 
&& =  3 \sum_{ B , C } \kappa_{ a BC } z_B z_C  = 3 \kappa_a ( z ) 
\eea
So we find 
\bea 
\kappa_a ( z ) = { 1 \over 3 } { \partial  \kappa ( z )  \over \partial z_a } 
\eea
Now use the result (\ref{resultKappa}) to find 
\bea 
\kappa_a ( z ) = a ( 1 - a^2 ) z_a^2 + \sum_{ A < a  } A ( 1 + A ) z_A^2 + \sum_{ a < B } 2 a ( 1 +a ) z_B z_a  
\eea 
Rewrite 
\bea
\kappa_A ( z ) = A ( 1 - A^2 ) z_A^2 + \sum_{ B < A } B ( 1 + B ) z_B^2 
+ \sum_{ A < B } 2 A ( 1 + A ) z_B z_A 
\eea

Using the above polynomial, we find the explicit form of the constraints to be 
\bea\label{ExpCons}  
&&  \hbox { For }  1 \le A \le (n-1)  : \cr 
&& \cr  
&& A ( 1 - A^2 ) X^{ (A) }_{ \mu} X^{ (A) }_{ \mu}  + \sum_{ B : B > A   } 2 A ( 1+A ) X^{(A)}_{\mu} X^{ (B)}_{ \mu} + \sum_{ B : B < A }  B  ( 1+ B  ) X^{(B)}_{\mu} X^{ (B)}_{ \mu} = 0 \cr 
&& \hbox{ and } \cr 
&& \cr 
&& \sum_{ A  =1}^{ n-1} X_{\mu}^{(A)}  X_{\mu}^{(A)}=0 
\eea

\section{ Examples of $ \cR / \cI$ at low $n,d$  } \label{conifold}

Recall that $V$ is the representation of $so(4,2)$ that has all the states which correspond, by the 
operator-state correspondence, to the fundamental field and its derivatives. 
The unrefined generating function for the fundamental field of $so(4,2)$ is
(the factor in front of the trace below removes the contribution from the dimension of the scalar field itself)
\bea 
s^{2-d\over 2}tr_V(s^D)={ ( 1 - s^2 ) \over ( 1- s)^4 } = {  (1 + s ) \over ( 1 - s )^3  }  
\eea
This is exactly  the unrefined Hilbert series for the conifold $\cC$ (see eq. (4.5) of \cite{BFHH06}. 
This is not an accident: the conifold is the solution set of 
\bea 
z_1^2  + z_2^2 + z_3^2 + z_4^2 = 0 
\eea
There are 4 generators and one quadratic relation, which matches the $so(4,2)$ problem. 
In constructing the basic irrep for the free scalar field, we look at states 
constructed by acting with $P_{\mu}$ on the ground state, and set to zero 
the equation of motion $P_{\mu}P_{\mu}=0$. 

For the case of the $d=4$,  $n=3$ primaries (skipping the $ n=2$ case where we have null vectors
to deal with), we have the unrefined counting function 
\bea
{(1-s^2)^3\over (1- s)^8} 
\eea 
This answer is intuitive: as we have explained, here we are looking at polynomials in 3 coordinates $x_{\mu}^1,x_{\mu}^2,x_{\mu}^3$. 
After solving the COM condition we have polynomials in the two coordinates $X_{\mu}^1, X_{\mu}^{2}$ which are annihilated by the Laplacian in $X^{1}$, the Laplacian in $X^{2}$, and by
\bea 
   \sum_{\mu=1}^d {\partial^2 \over \partial X_{ \mu}^{(1)} \partial X_{ \mu}^{(2)}  } 
\eea
This ring should be the ring of polynomials on $ \cC^{ 2} $ subject to the condition 
\bea 
\sum_{\mu=1}^d Z^{(1)}_{ \mu } Z^{(2)}_{ \mu} = 0 
\eea
So the counting of states in this case corresponds to a subvariety in the product of two copies of the conifold. 

In $d=3$ the ring with generators $z_1, z_2, z_3 $ with relation 
\bea 
z_1^2 + z_2^2 + z_3^2 = 0 
\eea
is the $A1$ singularity. 
So the $ n=3$ problem is a subvariety of the product of two $A1$ singularity.

It is easy to make further connections between the $so(3,2))$ counting of states in $V^{\otimes n}$ and 
the algebraic geometry of subvarieties.
These connections can be verified using the SAGE computer package.
For example, if $n=4$ we have the following four constraints (we have rescaled $X^{(3)}$ by $\sqrt{2}$)
\bea
X^{(1)}\cdot X^{(2)}+X^{(1)}\cdot X^{\prime (3)}&=&0\cr
X^{(1)}\cdot X^{(1)}-X^{(2)}X^{(2)}+X^{(2)}\cdot X^{\prime (3)}&=&0\cr
X^{(1)}\cdot X^{(1)}+X^{(2)}\cdot X^{(2)}-4X^{\prime (3)}\cdot X^{\prime (3)}&=&0\cr
X^{(1)}\cdot X^{(1)}+X^{(2)}\cdot X^{(2)}+2X^{\prime (3)}\cdot X^{\prime (3)}&=&0
\eea
Setting $d=3$ we have checked that we obtain the correct Hilbert series, by using the sage commands
\begin{verbatim}
R.<x1,x2,x3,y1,y2,y3,z1,z2,z3>=PolynomialRing(QQ,9); R
I=Ideal([
x1*y1+x2*y2+x3*y3+x1*z1+x2*z2+x3*z3,
x1^2+x2^2+x3^2-y1^2-y2^2-y3^2+y1*z1+y2*z2+y3*z3,
x1^2+x2^2+x3^2+y1^2+y2^2+y3^2-4*z1^2-4*z2^2-4*z3^2,
x1^2+x2^2+x3^2+y1^2+y2^2+y3^2+(z1^2+z2^2+z3^2)*2])
sage: I.hilbert_series()
\end{verbatim}
The Hilbert series we obtain is
\bea
H(s)={(1-s^2)^4\over (1-s)^9}
\eea
which is indeed correct.

For $n=5$, after a suitable rescaling, we have the following five constraints:
\bea
X^{(1)}\cdot X^{(2)}+X^{(1)}\cdot X^{(3)}+X^{(1)}\cdot X^{(4)} &=&0\cr
2 X^{(1)}\cdot X^{(1)}-6X^{(2)}X^{(2)}+12 X^{(2)}\cdot X^{(3)}+12 X^{(2)}\cdot X^{(4)}&=&0\cr
2 X^{(1)}\cdot X^{(1)}+6 X^{(2)}\cdot X^{(2)}-24 X^{(3)}\cdot X^{(3)}+24 X^{(4)}\cdot X^{(3)}&=&0\cr
2 X^{(1)}\cdot X^{(1)}+6 X^{(2)}\cdot X^{(2)}+12 X^{(3)}\cdot X^{(3)}-60 X^{(4)}\cdot X^{(4)}&=&0\cr
X^{(1)}\cdot X^{(1)}+ X^{(2)}\cdot X^{(2)}+ X^{(3)}\cdot X^{(3)}+ X^{(4)}\cdot X^{(4)}&=&0
\nonumber
\eea
Using the sage commands
\begin{verbatim}
R.<x1,x2,x3,y1,y2,y3,z1,z2,z3,w1,w2,w3>=PolynomialRing(QQ,12); R
I=Ideal([
x1*y1+x2*y2+x3*y3+x1*z1+x2*z2+x3*z3+x1*w1+x2*w2+x3*w3,
2*(x1^2+x2^2+x3^2)-6*(y1^2+y2^2+y3^2)+12*(y1*z1+y2*z2+y3*z3)
+12*(y1*w1+y2*w2+y3*w3),
2*(x1^2+x2^2+x3^2)+6*(y1^2+y2^2+y3^2)-24*(z1^2+z2^2+z3^2)
+24*(z1*w1+z2*w2+z3*w3),
2*(x1^2+x2^2+x3^2)+6*(y1^2+y2^2+y3^2)+12*(z1^2+z2^2+z3^2)
-60*(w1^2+w2^2+w3^2),
(x1^2+x2^2+x3^2)+(y1^2+y2^2+y3^2)+(z1^2+z2^2+z3^2)
+(w1^2+w2^2+w3^2)])
I.hilbert_series()\end{verbatim}
we again obtain the correct Hilbert series
\bea
H(s)={(1-s^2)^5\over (1-s)^{12}}
\eea

\section{ Derivation of a symmetric group multiplicity formula for $V_H^{\otimes k }$  }
\label{SnSkrefinedVHk} 

The multiplicity of $S_n \times S_k$ irreps in $V_H^{ \otimes k } $, where 
$V_{H} $ is the $S_n$ irrep  associated with Young diagram $[n-1,1]$, is obtained 
by taking the trace of appropriate  projectors which are expressible 
using characters \cite{FultonHarris}. We have
(see
for example \cite{BHR2} for further explanation of this formula)
\bea 
&& Dim ( V_{ \Lambda_2 , \Lambda_1 } ) = Mult ( V_H^{ \otimes k } , V_{ \Lambda_1}^{ S_n }  \otimes V_{\Lambda_2}^{ S_k }  ) \cr 
&& = { 1 \over n! k! } \sum_{ \sigma \in S_n } \sum_{ \tau \in S_k } \chi_{ \Lambda_1 } ( \sigma ) 
 \chi_{ \Lambda_2} ( \tau ) 
 \prod_{ i }  ( tr_H ( \sigma^i ) )^{ C_i ( \tau ) }  \cr 
  && =  \sum_{ p \vdash n } \sum_{ q \vdash k } 
  { \chi_{ \Lambda_1 }^p \chi^q_{ \Lambda_2} 
  \over Sym ~ p ~~~ Sym  ~ q }  \prod_{ i } ( tr_{ H } ( \sigma_p^i ) )^{ q_i } 
\eea
Here $ \sigma_p $ is a perm with cycle structure $p$. Now note that we have 
\bea 
tr_{V_H}  ( \sigma ) && = tr_{ nat } ( \sigma ) - tr_{ triv } ( \sigma ) \cr 
                && = C_1 ( \sigma ) - 1 
\eea
and 
\bea 
&& tr_H ( \sigma^i  ) =  C_1 ( \sigma^i  ) - 1  \cr 
&& = -1 + \sum_{ d | i  } d  C_d ( \sigma ) 
\eea
When we raise a permutation to power $i$, all cycles of length $d$ which divide $i$ 
split into $d$ cycles of length $1$. 
It follows 
\bea\label{refinedSnSkVH} 
&& Dim ( V_{ \Lambda_2 , \Lambda_1 } )
= \sum_{ p \vdash n } \sum_{ q \vdash k }{ \chi^p_{ \Lambda_1 }  \chi^q_{ \Lambda_2} 
  \over Sym ~ p ~~~ Sym  ~ q }  \prod_{i=1}^k  \left (    -1 + \sum_{ d | i  } d  p_d              \right )^{ q_i }
\eea

\section{ Refined counting of LWPs in $d=3$ dimensions : Tables }\label{d3count}

In this Appendix we summarize multiplicities for primaries constructed using $n$ fields and $k$ derivatives. 
These primaries transform in the spin $l$ representation of $so(3)$ and in the $\Lambda_n$ of $S_n$.

\begin{table}[H]
\center
\begin{tabular}{|c|c|c|}
\hline
$l$ & $\Lambda_3$ & Mult\\
\hline
2 & [3] & 1\\
\hline
2 & [2, 1]&  1\\
\hline
1 & [1, 1, 1] & 1\\
\hline
\end{tabular}
\caption{Results for $n=3$ fields and $k=2$ derivatives.}
\end{table}

\begin{table}[H]
\center
\begin{tabular}{|c|c|c|}
\hline
$l$ & $\Lambda_3$ & Mult\\
\hline
3 & [3] & 1\\
\hline
2 & [2, 1] & 1\\
\hline
3 & [2, 1] & 1\\
\hline
3 & [1, 1, 1] & 1\\
\hline
\end{tabular}
\caption{Results for $n=3$ fields and $k=3$ derivatives.}
\end{table}

\begin{table}[H]
\center
\begin{tabular}{|c|c|c|}
\hline
$l$ & $\Lambda_3$ & Mult\\
\hline
4 & [3] & 1\\
\hline
3 & [2, 1] & 1\\
\hline
4 & [2, 1] & 2\\
\hline
3 & [1, 1, 1] & 1\\
\hline
\end{tabular}
\caption{Results for $n=3$ fields and $k=4$ derivatives.}
\end{table}

\begin{table}[H]
\center
\begin{tabular}{|c|c|c|}
\hline
$l$ & $\Lambda_4$ & Mult\\
\hline
3 & [4] & 1\\
\hline
1 & [3, 1] & 1\\
\hline
2 & [3, 1] & 1\\
\hline
3 & [3, 1] & 2\\
\hline
2 & [2, 2] & 1\\
\hline
1 & [2, 1, 1] & 1\\
\hline
2 & [2, 1, 1] & 1\\
\hline
3 & [2, 1, 1] & 1\\
\hline
0 & [1, 1, 1, 1] & 1\\
\hline
\end{tabular}
\caption{Results for $n=4$ fields and $k=3$ derivatives.}
\end{table}

\begin{table}[H]
\center
\begin{tabular}{|c|c|c|}
\hline
$l$ & $\Lambda_5$ & Mult\\
\hline
3 & [5] & 1\\
\hline
1 & [4, 1] & 1\\
\hline
2 & [4, 1] & 1\\
\hline
3 & [4, 1] & 2\\
\hline
1 & [3, 2] & 1\\
\hline
2 & [3, 2] & 1\\
\hline
3 & [3, 2] & 1\\
\hline
1 & [3, 1, 1] & 1\\
\hline
2 & [3, 1, 1] & 1\\
\hline
3 & [3, 1, 1] & 1\\
\hline
1 & [2, 2, 1] & 1\\
\hline
2 & [2, 2, 1] & 1\\
\hline
0 & [2, 1, 1, 1] & 1\\
\hline
\end{tabular}
\caption{Results for $n=5$ fields and $k=3$ derivatives.}
\end{table}

\section{ Refined counting of LWPs in $d=4$ dimensions : Tables  }\label{d4count}

We give the multiplicities for primaries constructed using $n$ fields and $k$ derivatives. These primaries transform
in the spin $(l_1,l_2)$ representation of $so(4)$ and in the $\Lambda_n$ of $S_n$.

\begin{table}[H]
\center
\begin{tabular}{|c|c|c|c|}
\hline
$l_1$ & $l_2$ & $\Lambda_3$ & Mult\\
\hline
1 & 1 & [3] & 1\\
\hline
1 & 1 & [2, 1] & 1\\
\hline
0 & 1 & [1, 1, 1] & 1\\
\hline
1 & 0 & [1, 1, 1] & 1\\
\hline
\end{tabular}
\caption{Results for $n=3$ fields and $k=2$ derivatives.}
\end{table}

\begin{table}[H]
\center
\begin{tabular}{|c|c|c|c|}
\hline
$l_1$ & $l_2$ & $\Lambda_3$ & Mult\\
\hline
3/2 & 3/2 & [3] & 1\\
\hline
1/2 & 3/2 & [2, 1] & 1\\
\hline
3/2 & 1/2 & [2, 1] & 1\\
\hline
3/2 & 3/2 & [2, 1] & 1\\
\hline
3/2 & 3/2 & [1, 1, 1] & 1\\
\hline
\end{tabular}
\caption{Results for $n=3$ fields and $k=3$ derivatives.}
\end{table}

\begin{table}[H]
\center
\begin{tabular}{|c|c|c|c|}
\hline
$l_1$ & $l_2$ & $\Lambda_3$ & Mult\\
\hline
0 & 2 & [3] & 1\\
\hline
2 & 0 & [3] & 1\\
\hline
2 & 2 & [3] & 1\\
\hline
1 & 2 & [2, 1] & 1\\
\hline
2 & 1 & [2, 1] & 1\\
\hline
2 & 2 & [2, 1] & 2\\
\hline
1 & 2 & [1, 1, 1] & 1\\
\hline
2 & 1 & [1, 1, 1] & 1\\
\hline
\end{tabular}
\caption{Results for $n=3$ fields and $k=4$ derivatives.}
\end{table}

\begin{table}[H]
\center
\begin{tabular}{|c|c|c|c|}
\hline
$l_1$ & $l_2$ & $\Lambda_4$ & Mult\\
\hline
3/2 & 3/2 & [4] & 1\\
\hline
1/2 & 1/2 & [3, 1] & 1\\
\hline
1/2 & 3/2 & [3, 1] & 1\\
\hline
3/2 & 1/2 & [3, 1] & 1\\
\hline
3/2 & 3/2 & [3, 1] & 2\\
\hline
1/2 & 3/2 & [2, 2] & 1\\
\hline
3/2 & 1/2 & [2, 2] & 1\\
\hline
1/2 & 1/2 & [2, 1, 1] & 1\\
\hline
1/2 & 3/2 & [2, 1, 1] & 1\\
\hline
3/2 & 1/2 & [2, 1, 1] & 1\\
\hline
3/2 & 3/2 & [2, 1, 1] & 1\\
\hline
1/2 & 1/2 & [1, 1, 1, 1] & 1\\
\hline
\end{tabular}
\caption{Results for $n=4$ fields and $k=3$ derivatives.}
\end{table}

\begin{table}[H]
\center
\begin{tabular}{|c|c|c|c|}
\hline
$l_1$ & $l_2$ & $\Lambda_5$ & Mult\\
\hline
3/2 & 3/2 & [5] & 1\\
\hline
1/2 & 1/2 & [4, 1] & 1\\
\hline
1/2 & 3/2 & [4, 1] & 1\\
\hline
3/2 & 1/2 & [4, 1] & 1\\
\hline
3/2 & 3/2 & [4, 1] & 2\\
\hline
1/2 & 1/2 & [3, 2] & 1\\
\hline
1/2 & 3/2 & [3, 2] & 1\\
\hline
3/2 & 1/2 & [3, 2] & 1\\
\hline
3/2 & 3/2 & [3, 2] & 1\\
\hline
1/2 & 1/2 & [3, 1, 1] & 1\\
\hline
1/2 & 3/2 & [3, 1, 1] & 1\\
\hline
3/2 & 1/2 & [3, 1, 1] & 1\\
\hline
3/2 & 3/2 & [3, 1, 1] & 1\\
\hline
1/2 & 1/2 & [2, 2, 1] & 1\\
\hline
1/2 & 3/2 & [2, 2, 1] & 1\\
\hline
3/2 & 1/2 & [2, 2, 1] & 1\\
\hline
1/2 &1/2 &[2, 1, 1, 1] &1\\
\hline
\end{tabular}
\caption{Results for $n=5$ fields and $k=3$ derivatives.}
\end{table}

\end{appendix}


\begin{thebibliography}{} 

\bibitem{DRRR17-PRL}
  R.~de Mello Koch, P.~Rabambi, R.~Rabe and S.~Ramgoolam,
  Phys.\ Rev.\ Lett.\  {\bf 119} (2017) no.16,  161602
  doi:10.1103/PhysRevLett.119.161602
  [arXiv:1705.04039 [hep-th]].

\bibitem{DRRR17}
  R.~de Mello Koch, P.~Rabambi, R.~Rabe and S.~Ramgoolam,
  ``Counting and construction of holomorphic primary fields in free CFT4 from rings of functions on Calabi-Yau orbifolds,''
  JHEP {\bf 1708} (2017) 077
  doi:10.1007/JHEP08(2017)077
  [arXiv:1705.06702 [hep-th]].

\bibitem{HLMM17}
  B.~Henning, X.~Lu, T.~Melia and H.~Murayama,
  ``Operator bases, $S$-matrices, and their partition functions,''
  JHEP {\bf 1710} (2017) 199
  doi:10.1007/JHEP10(2017)199
  [arXiv:1706.08520 [hep-th]].

\bibitem{cft4tft2} 
  R.~de Mello Koch and S.~Ramgoolam,
  ``CFT$_4$ as $SO(4,2)$-invariant TFT$_2$,''
  Nucl.\ Phys.\ B {\bf 890} (2014) 302
  doi:10.1016/j.nuclphysb.2014.11.013
  [arXiv:1403.6646 [hep-th]].

\bibitem{Dolan0508}
  F.~A.~Dolan,
  J.\ Math.\ Phys.\  {\bf 47} (2006) 062303
  doi:10.1063/1.2196241
  [hep-th/0508031].
  
\bibitem{Cox}
D. A. Cox, J. B. Little and D. O'Shea, 
``Ideals, Varieties, and Algorithms'' Fourth Edition, 2015, Springer.

\bibitem{Eisenbud}
Eisenbud, David (1995), Commutative algebra. With a view toward algebraic geometry, Graduate Texts in Mathematics, 
150, New York: Springer-Verlag, ISBN 0-387-94268-8.

\bibitem{Grigorescu} 
E. Grigorescu, ``Hilbert series and free resolutions'' Bard College, senior thesis. 

\bibitem{WuGang}
Gang Wu, ``Koszul Algebras and Koszul Duality'' University of Ottawa, Masters Thesis. 

\bibitem{King} 
R C King, ``Young Tableaux, Schur functions and $ SU(2)$ plethysms'' J Phys A: Math Gen, 18, 2429. 

\bibitem{FultonHarris}
W. Fulton and J. Harris, ``Representation Theory: A First Course,'' Springer, 1991.

\bibitem{cayoub} 
C. Ayoub, ``On constructing bases for ideals in polynomial rings over the integers,'' 
Journal of number theory, 1983. 

 \bibitem{DRV18}
  R.~De Mello Koch, P.~Rambambi and H.~J.~R.~Van Zyl,
  ``From Spinning Primaries to Permutation Orbifolds,''
  JHEP {\bf 1804} (2018) 104
  doi:10.1007/JHEP04(2018)104
  [arXiv:1801.10313 [hep-th]].

\bibitem{RamChars} 
A. Ram, ``Characters of Brauer's Centralizer algebras'' Pacific Journal of Mathematics, 
Vol. 169, No. 1, 1985. 


\bibitem{cjr} 
  S.~Corley, A.~Jevicki and S.~Ramgoolam,
  ``Exact correlators of giant gravitons from dual N=4 SYM theory,''
  Adv.\ Theor.\ Math.\ Phys.\  {\bf 5} (2002) 809
  doi:10.4310/ATMP.2001.v5.n4.a6
  [hep-th/0111222].
  
\bibitem{QuivCalc} 
  J.~Pasukonis and S.~Ramgoolam,
  ``Quivers as Calculators: Counting, Correlators and Riemann Surfaces,''
  JHEP {\bf 1304} (2013) 094
  doi:10.1007/JHEP04(2013)094
  [arXiv:1301.1980 [hep-th]].

\bibitem{YusukeTFT2} 
  Y.~Kimura,
  ``Noncommutative Frobenius algebras and open-closed duality,''
  arXiv:1701.08382 [hep-th].
  
\bibitem{PCA} 
  P.~Mattioli and S.~Ramgoolam,
  ``Permutation Centralizer Algebras and Multi-Matrix Invariants,''
  Phys.\ Rev.\ D {\bf 93} (2016) no.6,  065040
  doi:10.1103/PhysRevD.93.065040
  [arXiv:1601.06086 [hep-th]].

  \bibitem{BCD08}
  R.~Bhattacharyya, S.~Collins and R.~de Mello Koch,
  ``Exact Multi-Matrix Correlators,''
  JHEP {\bf 0803} (2008) 044
  doi:10.1088/1126-6708/2008/03/044
  [arXiv:0801.2061 [hep-th]].

\bibitem{BDS08}
  R.~Bhattacharyya, R.~de Mello Koch and M.~Stephanou,
  ``Exact Multi-Restricted Schur Polynomial Correlators,''
  JHEP {\bf 0806} (2008) 101
  doi:10.1088/1126-6708/2008/06/101
  [arXiv:0805.3025 [hep-th]].


  \bibitem{KR1}
  Y.~Kimura and S.~Ramgoolam,
  ``Branes, anti-branes and brauer algebras in gauge-gravity duality,''
  JHEP {\bf 0711} (2007) 078
  doi:10.1088/1126-6708/2007/11/078
  [arXiv:0709.2158 [hep-th]].
  
  
  
 
  \bibitem{BHR1}
  T.~W.~Brown, P.~J.~Heslop and S.~Ramgoolam,
  ``Diagonal multi-matrix correlators and BPS operators in N=4 SYM,''
  JHEP {\bf 0802} (2008) 030
  doi:10.1088/1126-6708/2008/02/030
  [arXiv:0711.0176 [hep-th]].

\bibitem{BHR2} 
  T.~W.~Brown, P.~J.~Heslop and S.~Ramgoolam,
  ``Diagonal free field matrix correlators, global symmetries and giant gravitons,''
  JHEP {\bf 0904} (2009) 089
  [arXiv:0806.1911 [hep-th]].

\bibitem{PP} 
A. Polischchuk, L. Positselski, ``Quadratic Algebras'', University Lecture Series 37, Providence, R.I. 
American Mathematical Society.  

\bibitem{Wiki-Koszul} 
Wikipedia article on Koszul duality: https://en.wikipedia.org/wiki/Koszul\_duality

\bibitem{Szabo}
  R.~J.~Szabo,
  ``Quantum field theory on noncommutative spaces,''
  Phys.\ Rept.\  {\bf 378} (2003) 207
  doi:10.1016/S0370-1573(03)00059-0
  [hep-th/0109162].

\bibitem{FL2002} 
Froberg and Lofval, ``KOSZUL HOMOLOGY AND LIE ALGEBRAS WITH
APPLICATION TO GENERIC FORMS AND POINTS'', 
Homology, Homotopy and Applications, vol.4(2), 2002, pp.227–258

\bibitem{BFHH06} 
S.~Benvenuti, B.~Feng, A.~Hanany and Y.~H.~He,
  ``Counting BPS Operators in Gauge Theories: Quivers, Syzygies and Plethystics,''
  JHEP {\bf 0711}, 050 (2007)
  doi:10.1088/1126-6708/2007/11/050
  [hep-th/0608050].








  
\end{thebibliography}
\end{document}